\documentclass[journal]{IEEEtran}

\usepackage{cite}
\usepackage{amsmath,amssymb,amsfonts}
\usepackage{algorithmic}
\usepackage{graphicx}
\usepackage{setspace}
\usepackage{textcomp}
\usepackage{bm}
\usepackage{array}
\usepackage{url}
\usepackage{xcolor}
\usepackage{comment}
\usepackage{fancyhdr}
\newcommand{\1}{\mbox{1}\hspace{-0.25em}\mbox{l}}
\ifCLASSINFOpdf
\else
\fi
\newcommand{\MYheader}{\smash{\scriptsize
\hfil\parbox[t][\height][t]{\textwidth}{\centering
This paper has been published in IEEE Journal of Biomedical and Health Informatics (J-BHI).
Digital Object Identifier (DOI) is \protect\url{https://doi.org/10.1109/JBHI.2022.3227517}}\hfil\hbox{}}}

\newcommand{\MYfooter}{\smash{\scriptsize
\hfil\parbox[t][\height][t]{\textwidth}{\centering
© 2022 IEEE. Personal use of this material is permitted. Permission from IEEE must be obtained for all other uses, in any current or future media, including reprinting/republishing this material for advertising or promotional purposes, creating new collective works, for resale or redistribution to servers or lists, or reuse of any copyrighted component of this work in other works.}\hfil\hbox{}}}

\makeatletter

\def\ps@headings{%
\def\@oddhead{\mbox{}\scriptsize\MYheader \hfil \thepage}
\def\@evenhead{\scriptsize\thepage \hfil \MYheader\mbox{}}
\def\@oddfoot{\MYfooter}%
\def\@evenfoot{\MYfooter}}

\def\ps@IEEEtitlepagestyle{%
\def\@oddhead{\mbox{}\scriptsize\MYheader \hfil \thepage}%
\def\@evenhead{\scriptsize\thepage \hfil \MYheader\mbox{}}%
\def\@oddfoot{\MYfooter}%
\def\@evenfoot{\MYfooter}}

\makeatother

\begin{document}
%
\title{Edema Estimation from Facial Images Taken\\ Before and After Dialysis via Contrastive\\ Multi-Patient Pre-Training}
%
%
%

\author{Yusuke Akamatsu, Yoshifumi Onishi, Hitoshi Imaoka, Junko Kameyama, and Hideo Tsurushima 
\thanks{Yusuke Akamatsu, Yoshifumi Onishi, and Hitoshi Imaoka are with Biometrics Research Laboratories, NEC Corporation, Kawasaki, Japan (e-mail: yusuke-akamatsu@nec.com; y-onishi@nec.com; h-imaoka\_cb@nec.com). }
\thanks{Junko Kameyama was with the Department of Neurosurgery, Faculty of Medicine, University of Tsukuba, Tsukuba, Japan. She is 
now with the Institute of Medical Science, the University of Tokyo, Minato-ku, Japan (e-mail: Kameyama-junko@g.ecc.u-tokyo.ac.jp).}
\thanks{Hideo Tsurushima is with 
the Department of Neurosurgery, Faculty of Medicine, University of Tsukuba, Tsukuba, Japan (e-mail: hideo-tsurushima@md.tsukuba.ac.jp).}}

%
%

\markboth{Journal of \LaTeX\ Class Files,~Vol.~XX, No.~XX, December~2022}%
{Shell \MakeLowercase{\textit{et al.}}: Bare Demo of IEEEtran.cls for IEEE Journals}
%



\maketitle


\begin{abstract}
Edema is a common symptom of kidney disease, and quantitative measurement of edema is desired.
This paper presents a method to estimate the degree of edema from facial images taken before and after dialysis of renal failure patients.
As tasks to estimate the degree of edema, we perform pre- and post-dialysis classification and body weight prediction.
We develop a multi-patient pre-training framework for acquiring knowledge of edema and transfer the pre-trained model to a model for each patient.
For effective pre-training, we propose a novel contrastive representation learning, called weight-aware supervised momentum contrast (WeightSupMoCo).
WeightSupMoCo aims to make feature representations of facial images closer in similarity of patient weight when the pre- and post-dialysis labels are the same.
Experimental results show that our pre-training approach improves the accuracy of pre- and post-dialysis classification by 15.1\% and reduces the mean absolute error of weight prediction by 0.243 kg compared with training from scratch.
The proposed method accurately estimate the degree of edema from facial images; our edema estimation system could thus be beneficial to dialysis patients.
\end{abstract}


\section{Introduction}
\label{sec:introduction}
\IEEEPARstart{E}{dema} is the accumulation of excess fluid in tissues beyond the limits of physiological drainage~\cite{trayes2013edema}.
Edema can arise from a variety of causes, including kidney, central, circulatory, orthopedic, and metabolic diseases. 
Chronic accumulation, or more generalized edema, is due to the onset of chronic systemic conditions such as heart failure and kidney disease~\cite{ely2006approach}.
When the kidney is unable to remove water and waste products from the body due to renal failure, dialysis is performed to remove excess water from the blood.
At the end of 2018, there were approximately 550,000 dialysis patients in the United States~\cite{johansen2021us} and 340,000 in Japan~\cite{nitta20212018}.
It is crucial for dialysis patients to control weight gain from the appropriate body weight after dialysis~\cite{newmann2005adequacy,lindberg2007fluid}, which is called dry weight~\cite{thomson1967hemodialysis}.
Many studies have reported that large weight gain is associated with a higher risk of death~\cite{leggat1998noncompliance,foley2002blood,saran2003nonadherence,stegmayr2006minimized,movilli2007association}.
To control weight gain, dialysis patients need to limit fluid and salt intake in their daily lives.
It is expected that excessive water intake is associated with the presence of edema.
If the degree of edema can be estimated from the appearance of dialysis patients and the amount of fluid in the body can be predicted, it will assist in self-management of fluid intake on a daily basis.

A few studies have attempted to estimate the degree of edema by using images of edema areas~\cite{chen2018camera,smith2021objective}.
With the aim of estimating the degree of peripheral edema, Chen {\it et al.}~\cite{chen2018camera} captured images during a pitting test using a physical simulator that reproduced edema.
In the pitting test, the characteristics of peripheral edema are classified into four levels according to the degree of indentation when the edema area is compressed with a finger.
Subsequently, they estimated the edema level from the images by using a support vector machine (SVM)~\cite{cortes1995support} or convolutional neural network (CNN)~\cite{krizhevsky2012imagenet}.
Whereas Chen {\it et al.}~\cite{chen2018camera} attempted a camera-based edema level estimation by using a simulator, it is necessary to acquire those images during the pitting test.
Since the pitting test is usually performed by medical staff, it is difficult for patients to obtain images themselves for self-management of their fluid intake.
Also, they did not conduct a validation with actual patient images.
Smith {\it et al.}~\cite{smith2021objective} estimated edema levels of peripheral edema by using images of arms and limbs of healthy subjects and heart failure patients captured with short-wave infrared molecular chemical imaging (SWIR MCI).
By utilizing the large absorption coefficient of water in certain spectral regions of the SWIR MCI, a regression model was constructed to estimate the edema level from the spectra.
Unlike Chen {\it et al.}~\cite{chen2018camera}, Smith {\it et al.}~\cite{smith2021objective} used data from actual heart failure patients, but required a special SWIR camera.
In daily fluid volume management for dialysis patients, it is difficult to use a special camera at home or outside the home.
Therefore, a simpler method to estimate the degree of edema is desired.

This study attempts to estimate the degree of edema by using facial images of dialysis patients captured with a tablet device, which is relatively easy for patients to use.
Compared with images of arms and legs used in previous studies, facial images can be taken more easily by patients themselves.
Also, diagnosis using facial images could be easily incorporated in video-based telemedicine.
Furthermore, by combining facial recognition systems, it becomes possible to acquire facial images as soon as the patient arrives at the hospital.
To the best of our knowledge, this is the first study to estimate edema from facial images of dialysis patients captured with a common visible-light camera.

We use facial images taken before and after dialysis of renal failure patients to estimate the degree of edema in two ways: i) pre- and post-dialysis classification and ii) weight prediction.
In the pre- and post-dialysis classification, we classify the facial images into two categories: pre-dialysis, when edema is present, and post-dialysis, when edema is alleviated.
In the weight prediction, we estimate the degree of edema by predicting the body weight, which reflects the fluid volume in the dialysis patient's body.
Although body weight changes may be due to alterations in activity, diet, and medication use~\cite{webel2007daily}, body weight is usually recommended as a proxy measurement of edema~\cite{chausiaux2021evaluation}.
In particular, the interdialytic weight gain is a function of oral fluid intake, which includes routine food ingestion~\cite{kalantar2009fluid}.
Since fluid overload usually implies the degree of edema~\cite{granado2016fluid}, we can assume that the degree of edema is correlated with body weight in dialysis patients.
The pre- and post-dialysis classification is a straightforward evaluation to identify facial changes before and after dialysis, and the weight prediction is a more practical task to evaluate our method.
By using pre- and post-dialysis labels and body weight as ground-truth, it becomes unnecessary to perform labeling tasks such as like having medical staff subjectively annotate edema levels, as was done in the previous study~\cite{smith2021objective}.

\begin{figure}[t]
\begin{center}
\includegraphics[scale=0.47]{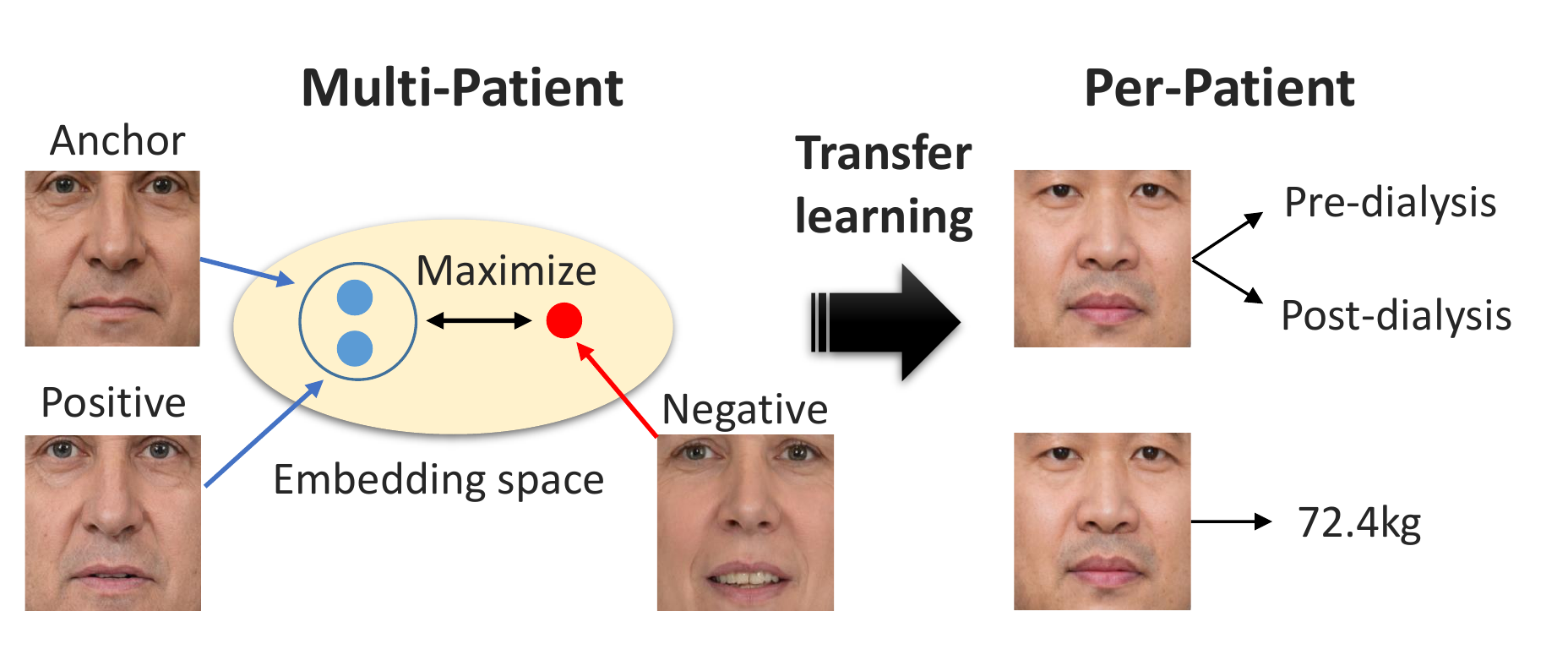}
\end{center}
\vspace{-10pt}
\caption{Our edema estimation framework. First, we use multi-patient data to pre-train the model to acquire knowledge about edema. Then, we perform pre- and post-dialysis classification and weight prediction by transfer learning on per-patient data. Note that the test patient in per-patient data is different from the patients in the multi-patient data used for pre-training.}
\label{fig:intro}
\end{figure}

As a first attempt, we trained a single CNN-based model across all patients for the pre- and post-dialysis classification.
However, the classification accuracy for new patients not in the training data reached only 59.7\% (see Results section~\ref{subsec:experiment-results}).
This low level of performance may have been because the changes in the face before and after dialysis are subtle and also because the faces vary widely from patient to patient.
Therefore, this study focuses on training a model for each patient to identify subtle changes in the face before and after dialysis.
In this situation, the training data for the model is limited to a small dataset obtained from individual patients.
To overcome this limitation, we pre-train the model using multi-patient data.
Specifically, we first pre-train the model to acquire knowledge about edema from multi-patient data and then perform transfer learning using per-patient data.

In this paper, we propose a novel contrastive learning method for pre-training, called {\it weight-aware supervised momentum contrast (WeightSupMoCo)}.
Contrastive learning is self-supervised or supervised representation learning that differentiates between similar example pairs and dissimilar example pairs~\cite{le2020contrastive,liu2021self,jaiswal2020survey,shurrab2022self}.
WeightSupMoCo is a supervised representation learning that leverages pre- and post-dialysis labels and patient weight.
The aim of the WeightSupMoCo loss is to make feature representations of facial images closer in similarity of patient weight when the pre- and post-dialysis labels are the same.
First, contrastive multi-patient pre-training is performed based on WeightSupMoCo and then pre- and post-dialysis classification and weight prediction are performed by transfer learning on per-patient data (see Fig.~\ref{fig:intro}~\footnote{For illustrative purposes, this paper does not use facial images obtained from actual patients, but rather facial images from Generated Photos (\url{https://generated.photos/}).}).
Experiments using data from 39 dialysis patients demonstrate that the contrastive multi-patient pre-training based on WeightSupMoCo significantly improves edema estimation performance.
Our main contributions are summarized as follows:
\begin{itemize}
\item This is the first work to estimate the degree of edema from facial images taken before and after dialysis. 
The main purpose of this study is to verify the feasibility of the estimation and to develop a deep learning model suitable for this purpose.
\item We construct an edema estimation model for each patient after the pre-training on multi-patient data.
We propose a novel contrastive multi-patient pre-training method to acquire knowledge about edema.
\item We validate the edema estimation by using facial images collected from 39 patients on a total of 210 dialysis days.
Our edema estimation system has the potential to become a health management tool for dialysis patients.
\end{itemize}
The rest of this paper is organized as follows:
Section~\ref{sec:related} reviews the related work, and section~\ref{sec:method} explains the proposed method.
Section~\ref{sec:experiments} reports experimental results, and section~\ref{sec:discussions} discusses our edema estimation system.
Section~\ref{sec:conclusions} provides the conclusions of this paper.

\section{Related Work}
\label{sec:related}
This section reviews recent work in three key areas related to this paper: edema measurement methods (\ref{subsec:edema}), disease identification from facial images (\ref{subsec:disease}), and contrastive representation learning (\ref{subsec:contrastive}).

\begin{figure*}[t]
\begin{center}
\includegraphics[scale=0.4]{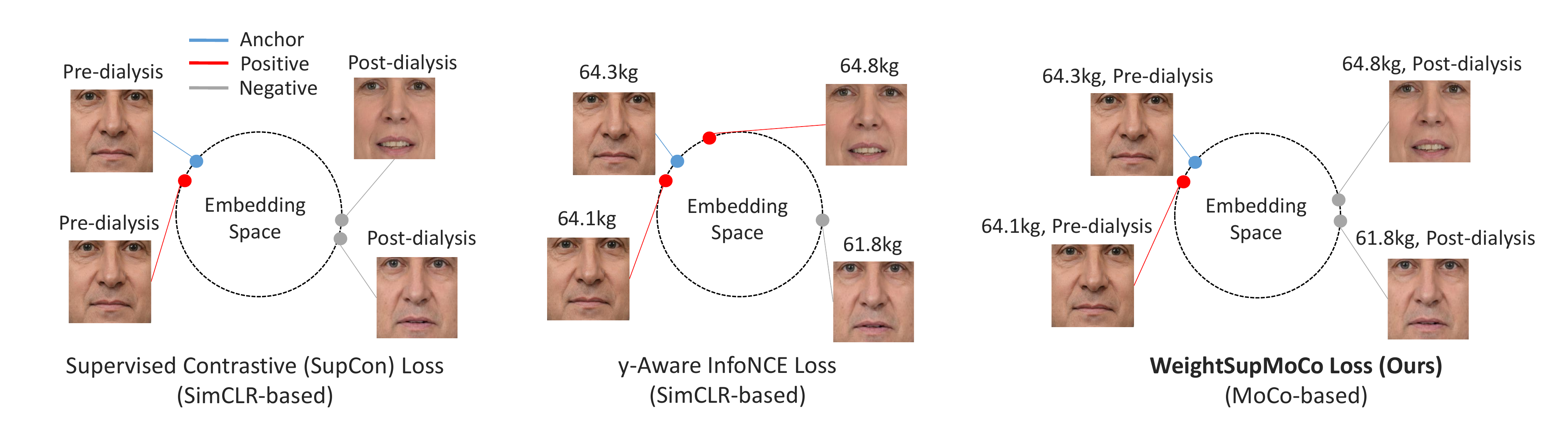}
\end{center}
\vspace{-10pt}
\caption{Differences between Supervised Contrastive (SupCon), {\it y}-Aware InfoNCE, and WeightSupMoCo (Ours) losses in multi-patient pre-training for edema estimation. The SupCon and {\it y}-Aware InfoNCE losses use discrete ({\it i.e.}, pre- and post-dialysis labels) and continuous ({\it i.e.}, patient weight) labels, respectively. On the other hand, the WeightSupMoCo loss uses both discrete and continuous labels to make the feature representations of the facial images closer in similarity of patient weight when the pre- and post-dialysis labels are the same. Moreover, SupCon and {\it y}-Aware InfoNCE are SimCLR-based contrastive learning methods, whereas WeightSupMoCo is a MoCo-based contrastive learning method.}
\label{fig:contrastive}
\end{figure*}

\subsection{Edema Measurement Methods}
\label{subsec:edema}
Traditional methods for measuring peripheral edema use a pitting test~\cite{raju2002clinical} or a medical tape to measure ankle circumference~\cite{mayrovitz2003limb}. 
Both of these methods have significant intra- and inter-measurer variability~\cite{brodovicz2009reliability} and require a medical staff to take the measurements. 
Another traditional method is measuring limb volume using water displacement~\cite{casley1994measuring}, which can measure the total foot volume and thus edema. 
Although water displacement volumetry is accurate and  sensitive~\cite{brodovicz2009reliability}, it takes a lot of time (25 minutes per foot~\cite{chausiaux2021evaluation}) and labor for the patient. 
Some recent work~\cite{chausiaux2021evaluation,mestre2014validation,kiyomitsu2017volume,masui2018visualization} has utilized 3D imaging to measure foot volume as alternative technologies to water displacement. 
However, they require expensive scanners~\cite{mestre2014validation}, a 3D camera installed on the wall~\cite{chausiaux2021evaluation}, and scanning around the feet with a hand-held depth camera~\cite{kiyomitsu2017volume,masui2018visualization}, which are still difficult for self-management of fluid intake.

To address these issues, we aim to develop a method that helps dialysis patients easily self-manage their fluid intake. 
We focus on facial images taken before and after dialysis and attempt to estimate the degree of edema from the facial images. 
Facial images can be captured more easily by patients themselves than foot images. 
They can also use a tablet or smartphone to capture facial images without introducing additional special equipment.
Furthermore, our image-based method can be incorporated into telemedicine, which is becoming more widespread these days.

\vspace{-5pt}
\color{black}
\subsection{Disease Identification from Facial Images}
\label{subsec:disease}
In recent years, several studies have identified developmental disorders~\cite{shukla2017deep}, genetic disorders~\cite{gurovich2019identifying}, acute illness~\cite{forte2021deep}, and Alzheimer’s disease~\cite{umeda2021screening} from facial images.
They employed deep learning approaches to identify diseases from facial images.
Gurovich {\it et al.}~\cite{gurovich2019identifying} predicted hundreds of genetic syndromes from facial images by fine-tuning the CNN model pre-trained on a large-scale face identity database~\cite{yi2014learning}.
Forte {\it et al.}~\cite{forte2021deep} trained a CNN model to classify facial images into healthy or acutely ill by using sets of simulated sick faces and synthetically generated data. 
These disease identifications recognize facial features such as facial dysmorphism~\cite{shukla2017deep,gurovich2019identifying}, specific symptoms appearing on a part of the face~\cite{forte2021deep}, or a specific complexion~\cite{umeda2021screening}.
A common CNN model across all patients was constructed in these studies.

On the other hand, edema appears and decreases subtly on each patient's face.
Therefore, it is difficult for a common model across all patients to capture such subtle changes in each patient.
Here, we construct a CNN model for each individual patient, rather than a common model across all patients.
We pre-train a model to acquire feature representations of edema from multi-patient data before fine-tuning it to each patient.

\begin{figure*}[t]
\begin{center}
\includegraphics[scale=0.45]{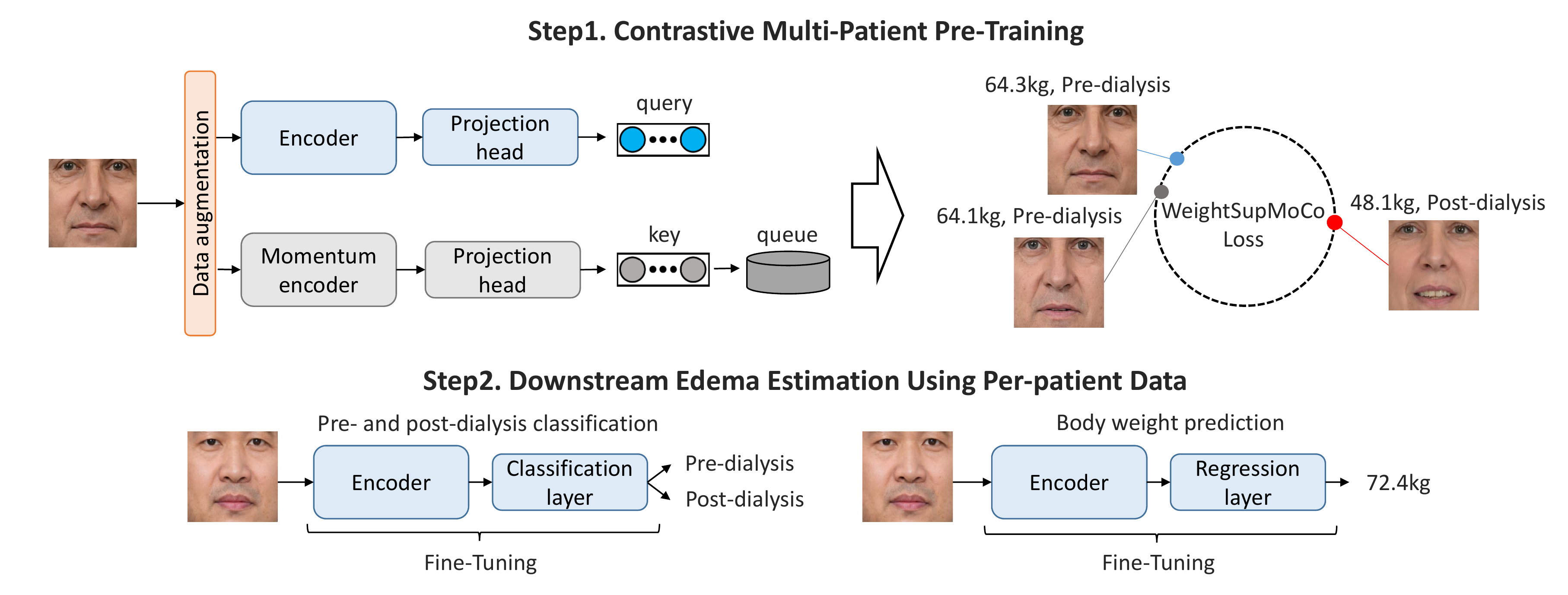}
\end{center}
\vspace{-10pt}
\caption{Overview of the proposed method. In Step 1, contrastive multi-patient pre-training based on WeightSupMoCo is performed. In Step 2, the encoder pre-trained in Step 1 is transferred to the downstream edema estimation tasks by fine-tuning using per-patient data. The downstream edema estimation tasks consist of pre- and post-dialysis classification and weight prediction.}
\label{fig:overview}
\end{figure*}

\subsection{Contrastive Representation Learning}
\label{subsec:contrastive}
Contrastive learning has recently become the mainstream of self-supervised~\cite{van2018representation,chen2020simple,he2020momentum,grill2020bootstrap,caron2020unsupervised} or supervised~\cite{khosla2020supervised,dufumier2021contrastive} representation learning.
In order to reduce manual annotations, self-supervised learning approaches produce {\it pseudo-labels} by applying specific augmentations or transformations to the input data.
Subsequently, the model pre-trained or trained with the augmented/transformed data and associated {\it pseudo-labels} can be used in downstream tasks, {\it e.g.}, classification and regression.
A simple framework for contrastive learning of visual representations (SimCLR)~\cite{chen2020simple} is one of the most popular self-supervised contrastive learning approaches.
SimCLR first creates a pair of positively correlated views by applying augmentations including cropping, flipping, and color distortion to the input image.
Then, both views are passed into a pair of convolutional encoders ({\it e.g.}, ResNet50~\cite{he2016deep}) and projection heads ({\it e.g.}, two dense layers with ReLU activation~\cite{nair2010rectified}) to obtain their embedding vectors.
SimCLR optimizes the whole architecture by maximizing the agreement between the positive pair of embedding vectors ({\it i.e.}, augmented views of the same image) while minimizing the negative pairs ({\it i.e.}, different images) in the same mini-batch.
Finally, the projection heads are discarded while the convolutional encoders are kept to be utilized in downstream tasks.
Momentum contrast (MoCo)~\cite{he2020momentum} is another popular self-supervised contrastive learning approach.
Instead of a pair of encoders as in SimCLR, MoCo uses an encoder and momentum encoder pair.
The momentum encoder generates a dictionary as a queue of encoded keys by enqueuing the current mini-batch and dequeuing the oldest mini-batch.
The advantage of using the momentum encoder is that the dictionary size is not restricted to the mini-batch size and can be larger.
Whereas the number of negative samples available in SimCLR is limited to the mini-batch size, MoCo can cover a rich set of negative samples and learn better feature representations.
Contrastive learning has been extended to supervised learning that makes use of the output labels~\cite{khosla2020supervised,dufumier2021contrastive}.
In supervised contrastive (SupCon) learning~\cite{khosla2020supervised}, the positive samples are generated not only from augmentations of the same input, but also from augmented views of other samples in the {\it same class}.
While SupCon assumes the use of {\it discrete} labels as the output labels, contrastive learning utilizing {\it continuous} labels has also been proposed~\cite{dufumier2021contrastive}.
Dufumier {\it et al.}~\cite{dufumier2021contrastive} introduced the {\it y}-Aware InfoNCE loss for 3D brain MRI classification to make feature representations of MRI images closer in similarity of patient age.
The utilization of {\it continuous} labels ({\it e.g.}, patient age) helps the model to learn a more generalizable representation of the data.

Our WeightSupMoCo loss leverages both {\it discrete} and {\it continuous} labels, {\it i.e.}, pre- and post-dialysis labels and patient weight.
Both of these two label types have important information about edema. 
Specifically, the pre- and post-dialysis labels provide coarse-grained information about edema while patient weight provides fine-grained information.
Therefore, these two label types lead to better representation learning in the multi-patient pre-training step.
Furthermore, we employ MoCo-based contrastive learning, which allows us to cover a richer set of positive and negative samples compared with the SimCLR-based SupCon~\cite{khosla2020supervised} and {\it y}-Aware InfoNCE~\cite{dufumier2021contrastive} losses.
Figure~\ref{fig:contrastive} illustrates the differences between the SupCon, {\it y}-Aware InfoNCE, and WeightSupMoCo losses in our pre-training setting.

\begin{figure}[t]
\begin{center}
\includegraphics[scale=0.35]{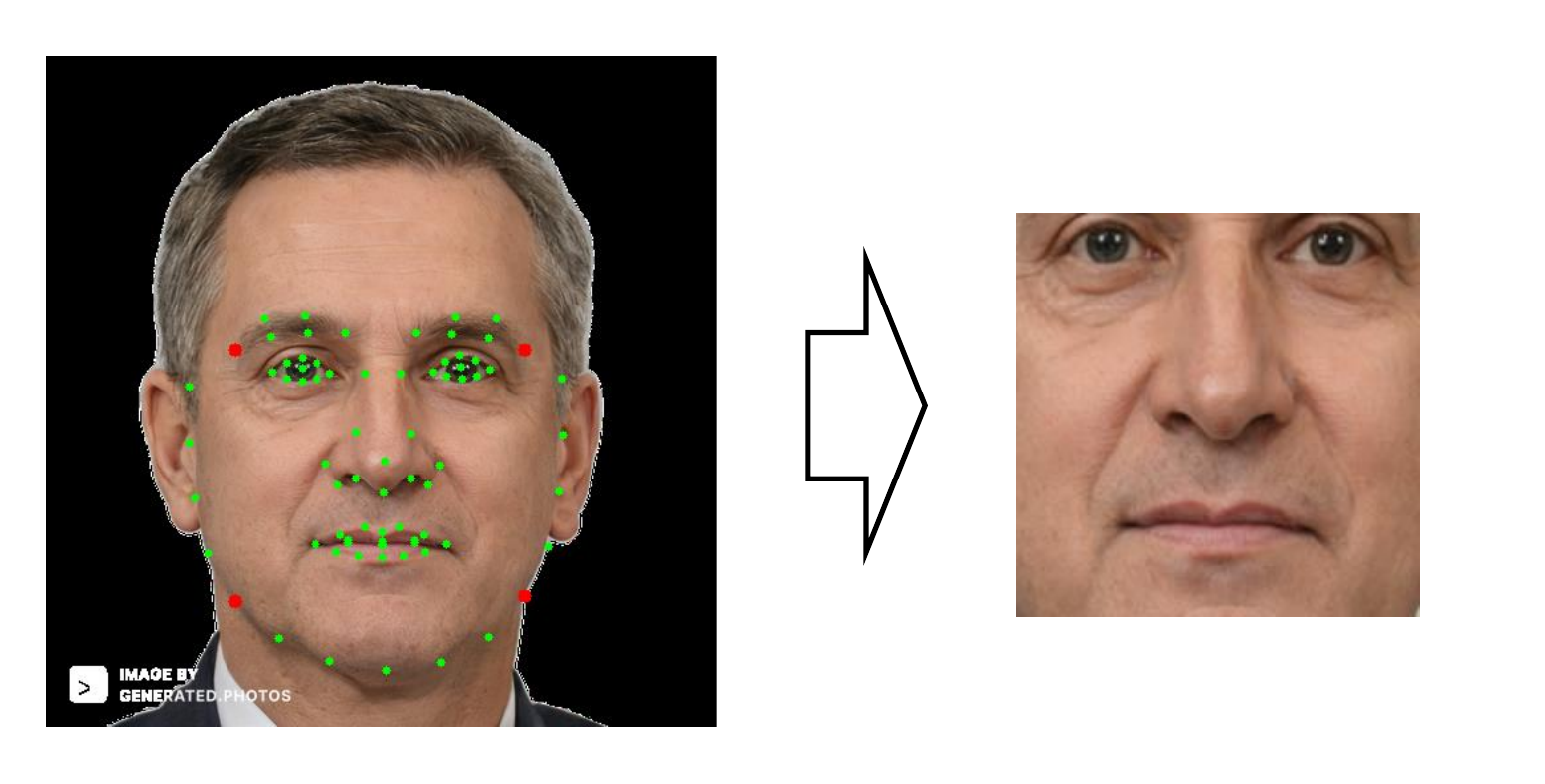}
\end{center}
\vspace{-10pt}
\caption{Face alignment and extraction of the center of the face.}
\label{fig:face}
\vspace{-10pt}
\end{figure}

\section{Method}
\label{sec:method}
All procedures in this study were approved by the Ethical Review Board at University of Tsukuba (the protocol number: 1663, the date of approval: August 16, 2021).
Figure~\ref{fig:overview} shows an overview of the proposed method.
First, we perform contrastive multi-patient pre-training based on WeightSupMoCo.
Then, we perform pre- and post-dialysis classification and weight prediction by transfer learning on per-patient data.
In \ref{subsec:method-contrastive}, we describe the main components of our contrastive multi-patient pre-training framework.
The WeightSupMoCo loss is detailed in \ref{subsec:method-weightsupmoco} and the downstream edema estimation using per-patient data is presented in \ref{subsec:method-downstream}.

\subsection{Contrastive Multi-patient Pre-training Components}
\label{subsec:method-contrastive}
First, we obtain facial images of renal failure patients taken before and after dialysis.
Then, we apply a facial detection method~\cite{imaoka2021future} to the obtained facial images and extract the centers of the facial images. 
Specifically, as shown in Fig.~\ref{fig:face}, we detect facial landmarks (green points) and extract the center part of the face by using four red points.
By detecting facial landmarks for each patient and extracting the center part based on these landmarks, we compensate for rotation due to head or body tilt. 
Also, to compensate for the facial size, all extracted images are scaled to the same size.
The center of the face includes the eyes and nose, where edema may appear and does not include areas unrelated to edema such as hair or clothes.
Subsequently, the facial images are used in our MoCo-based contrastive learning framework (see Fig.~\ref{fig:overview}. Step1).
MoCo generates a {\it query} and {\it key} from the input images via the encoder and momentum encoder, respectively.
Then, the encoder is trained by matching an encoded query to a dictionary of encoded keys (called a {\it queue}) by using the WeightSupMoCo loss (see \ref{subsec:method-weightsupmoco}).
The main components of our framework are the following data augmentation module, encoder network, and projection head:

\textbf{Data Augmentation Module,} $Aug(\cdot)$. Various combinations of data augmentations are randomly applied to each input image $\bm{x}$, as $\bm{\hat{x}} = Aug(\bm{x})$.
The input images $\bm{x}$ are resized to 224$\times$224.
The data augmentations consist of random horizontal flip ($p=0.5$), random color jitter ($p=0.8$, brightness $=0.4$, contrast $=0.4$, saturation $=0.4$, hue $=0.1$), and random grayscale conversion ($p=0.2$), where $p$ denotes a probability.
These data augmentation settings are derived from the previous work\cite{he2020momentum,wu2018unsupervised}.
We generate two augmented views of each input image $\bm{x}$.

\textbf{Encoder Network,} $Enc(\cdot)$, is used for both the encoder and the momentum encoder.
The encoder network maps $\bm{\hat{x}}$ to a representation vector, $\bm{r} = Enc(\bm{\hat{x}}) \in \mathbb{R}^{D_r}$.
The augmented images are separately input to the encoder and the momentum encoder, resulting in a pair of representation vectors.
In our experiment, we use ResNet18~\cite{he2016deep} as the encoder network; the dimension of representation vectors $D_r$ is 512.
Previous studies~\cite{chen2020simple,he2020momentum,khosla2020supervised,dufumier2021contrastive} have shown that encoder pre-trained in contrastive learning can be used for several related downstream tasks. 
As in the previous studies, we perform transfer learning by adding a linear layer for each downstream task to the same encoder.

\textbf{Projection Head,} $Proj(\cdot)$, maps the representation vector $\bm{r}$ onto an embedding vector, $\bm{z} = Proj(\bm{r}) \in \mathbb{R}^{D_z}$.
As in the previous work~\cite{chen2020simple,khosla2020supervised}, we instantiate $Proj(\cdot)$ as two dense layers with ReLU activation.
The output dimension of the first layer is 512 and that of the second layer $D_z$ is 128.
We normalize the output of the second layer to lie on the unit hypersphere, so that inner products can be used to measure cosine similarities in the embedding space.
Although the projection head is an important component to project feature representations onto the embedding space, it is not required for the downstream edema estimation tasks.
Therefore, we discard the projection head after the contrastive multi-patient pre-training and use only the encoder for the downstream tasks.

\subsection{WeightSupMoCo Loss}
\label{subsec:method-weightsupmoco}
The embedding vector $\bm{z}$ is used to optimize the encoder and projection head.
The common loss function for self-supervised contrastive learning in SimCLR and MoCo is the following InfoNCE loss~\cite{van2018representation}:
\begin{flalign}
    \mathcal{L}_i^{self} = - \log \frac{\exp(\bm{z}_{i}\cdot \bm{z}_{p} /\tau)}{\sum_{j\in A(i)} \exp(\bm{z}_{i}\cdot \bm{z}_{j} /\tau)},
\end{flalign}
where index $i$ is called the {\it anchor} and index $p$ is called the {\it positive} ({\it i.e.}, the augmented view of the anchor).
Here, $A(i)$ denotes the indices of all augmented samples except anchor $i$, consisting of one positive $p$ and all other {\it negatives} ({\it i.e.}, different images from the anchor).
The $\cdot$ symbol denotes the inner product and $\tau$ is a temperature parameter.

The WeightSupMoCo loss aims to make the embedding vectors closer in similarity of patient weight when the pre- and post-dialysis labels are the same.
This idea is based on two assumptions: (i) the pre- and post-dialysis labels indicate the presence or absence of edema, and (ii) patient weight indicates the degree of edema.
These two label types thus provide coarse- and fine-grained information about edema.
Let $l$ denote the pre- or post-dialysis label and $y$ denote the patient weight; the WeightSupMoCo loss is defined as follows:
\begin{flalign}
    \mathcal{L}_i^{ours} = - \sum_{k\in A(i)} & \frac{w_{\sigma}(y_i,y_k)\cdot \1_{l_i=l_k}}
    {\sum_{j\in A(i)} w_{\sigma}(y_i,y_j)\cdot \1_{l_i=l_j}} \nonumber \\
    &\quad\quad \times \log \frac{\exp(\bm{z}_{i}\cdot \bm{z}_{k} /\tau)}{\sum_{j\in A(i)} \exp(\bm{z}_{i}\cdot \bm{z}_{j} /\tau)},
\end{flalign}
where $w_{\sigma}(\cdot,\cdot)$ is a function representing the similarity of patient weights, in which we use the radial basis function (RBF) kernel, as in the {\it y}-Aware InfoNCE loss~\cite{dufumier2021contrastive}.
$\1_{i=j} \in \{0,1\}$ is an indicator function, which outputs 1 when $i=j$, and 0 otherwise.
The hyperparameter of the RBF kernel $\sigma$ is set to 3.0 in our case, and the temperature parameter $\tau$ is set to 0.1 as in ~\cite{chen2020simple,dufumier2021contrastive}.
In Eq. (2), when the pre- and post-dialysis label is the same and the similarity of weights is large, the value of $w_{\sigma}(y_i,y_k)\cdot \1_{l_i=l_k}$ increases and thus the embedding vectors are trained to be closer.
On the other hand, when the pre- and post-dialysis label is different, $\1_{l_i=l_k}$ becomes 0 and thus the distance of the embedding vectors is maximized.
In our MoCo-based framework, the anchor $i$ is the set of all query and queue samples, and $A(i)$ is the set of query samples except anchor $i$ in the same mini-batch and queue samples.
The dictionary size of the queue (1024 in our case) can be much larger than the mini-batch size of the query and key (16 in our case).
In the queue, the encoded key samples of the current mini-batch are enqueued and those of the oldest mini-batch are dequeued.
Hence, the queue can hold newer keys and remove older ones.

Although the use of queues enables the dictionary to be large, updating the momentum encoder by back-propagation is intractable because the gradient should propagate to all samples in the queue.
This issue can be addressed by using the following momentum update~\cite{he2020momentum}:
\begin{equation}
    \theta_k \gets m \theta_k + (1-m) \theta_q,
\end{equation}
where $\theta_q$ and $\theta_k$ respectively denote the parameters of the encoder and momentum encoder, and $m \in [0,1)$ is the momentum coefficient.
Only the $\theta_q$ is updated by back-propagation, while the $\theta_k$ evolves more smoothly than $\theta_q$ through the momentum update.
Following the suggestion in \cite{he2020momentum}, the momentum coefficient is set relatively large (0.9999 in our case) so that the $\theta_k$ will evolve slowly.

\subsection{Downstream Edema Estimation}
\label{subsec:method-downstream}
After the contrastive multi-patient pre-training based on WeightSupMoCo, we perform the downstream edema estimation tasks by transfer learning on per-patient data (see Fig.~\ref{fig:overview}. Step2).
The downstream edema estimation tasks comprise pre- and post-dialysis classification and weight prediction.
In these tasks, we discard the projection head and add a linear layer to the pre-trained encoder.
The projection head is used to project the embedding vector in contrastive learning, while the linear layer is used for the pre- and post-dialysis classification and weight prediction in downstream tasks, as in~\cite{khosla2020supervised,dufumier2021contrastive}.
As the linear layer, we use the classification layer with two output dimensions for the pre- and post-dialysis classification.
We also use the regression layer with one output dimension for the weight prediction.
Finally, the entire model including the encoder and linear layer is fine-tuned with per-patient data.
As the loss functions for the fine-tuning, we use the cross-entropy loss for the pre- and post-dialysis classification and the mean squared error (MSE) loss for the weight prediction.

\begin{table}[t]
\caption{Characteristics of the dataset (top) and dialysis patients (bottom). Means and standard deviations, except for gender, are shown at the bottom.}
\vspace{-5pt}
\label{table:dataset}
\begin{center}
\scalebox{0.9}{
\begin{tabular}{l|l}  \hline
Number of patients & 39 \\
Total dialysis days & 210 \\
Number of pre-dialysis data & 210 \\ 
Number of post-dialysis data & 182 \\ 
Total number of data & 392 \\
Total number of facial images & 39,200 \\
Average number of images per patient & 1,005 \\ \hline
Age (years) & 73.5 $\pm$ 9.4\\
Gender (male/female) & 25 / 14\\
Length of time on dialysis (months) & 81.0 $\pm$ 54.8\\
Weight before dialysis (kg) & 59.9 $\pm$ 11.6\\
Weight after dialysis (kg) & 57.8 $\pm$ 11.5\\
Water removal (ml) & 2214.0 $\pm$ 639.7\\ \hline
\end{tabular}
}
\end{center}
\end{table}

\section{Experiments}
\label{sec:experiments}

\subsection{Dataset}
\label{subsec:experiment-dataset}
To collect facial images from patients before and after dialysis, University of Tsukuba recruited dialysis patients in the Takemura Medical Nephro Clinic, Japan.
We explained the purpose, methods, and freedoms in participating in the study to each patient and obtained consent from all patients.

A total of 40 patients participated in the data collection, and data ranging from 1 to 10 dialysis days were obtained for each patient.
The number of participation days depended on the patient's freedom.
These dialysis patients have renal failures such as diabetic nephropathy and chronic glomerulonephritis.
On each dialysis day, we acquired data twice, before and after dialysis.
Data were not obtained for several days after dialysis due to the patient's physical condition.
Facial videos and medical records including age, body weight (before and after dialysis), and water removal were collected from each patient.
One patient was excluded from the dataset because a valid medical record could not be obtained; hence, data from 39 patients were used in the experiment.
The characteristics of the dataset and dialysis patients are listed in Table~\ref{table:dataset}.
The relationship between the number of days and the number of patients is shown in Appendix~A.

Facial videos were recorded using the built-in camera of a Surface Pro 7~\footnote{\url{https://www.microsoft.com/en-us/surface/}}, a popular tablet device.
Three minutes of video was captured, including frontal face and side views from the left and right while the patient was performing several tasks ({\it e.g.}, rest, opening mouth, raising hands).
In this experiment, we used only the facial videos from the frontal view at rest.
Then, we randomly selected arbitrary frames from the video and obtained 100 images from each video, resulting in a total of 39,200 facial images.
We excluded facial images when looking left or right based on the distance between the eye and nose landmarks (see Fig~\ref{fig:face}). 
We also excluded facial images when opening the mouth ({\it e.g.}, while talking) based on the distance between the mouth landmarks (see Fig~\ref{fig:face}).
We recorded the facial videos indoors with no direct sunlight. 
However, since lighting conditions could still vary from day to day, we apply data augmentation that changes the color of the image ({\it i.e.}, color jitter and grayscale conversion). 
By changing the color of the image, we can mitigate over-fitting to lighting conditions and improve the performance of edema estimation (see Appendix C).
The data of the 39 patients were divided into a contrastive multi-patient pre-training set (see Fig.~\ref{fig:overview}, Step1) and downstream edema estimation set (see Fig.~\ref{fig:overview}, Step2).
In this experiment, data from 24 patients with less than 7 days of pre- and post-dialysis were used for the pre-training, and data from 15 patients with 7 or more days were used for the downstream edema estimation (see Appendix A).

\subsection{Experimental Settings}
\label{subsec:experiment-settings}
For the multi-patient pre-training, 80\% of the images were randomly selected as training data and the remaining 20\% were used as validation data.
For the downstream edema estimation, we performed a {\it leave-one-day-out cross-validation}, where the data from one day were used as the test data and the data from the other days were used as the training and validation data.
The training and validation data ratios in this cross-validation were 80\% and 20\%, respectively.
The method of splitting training and validation data in the multi-patient pre-training is discussed in Appendix B.

The framework was implemented with PyTorch~\cite{paszke2019pytorch} on an NVIDIA GeForce RTX 3060 GPU.
During training, we used the Adam optimizer~\cite{kingma2015adam} with a learning rate of 1e-4 and weight decay of 5e-5 for the pre-training and a learning rate of 1e-3 for the downstream tasks.
We trained for 100 epochs in the pre-training and 20 epochs in the downstream tasks. 
The best models using the validation data were used in the pre-training and downstream tasks.
As in the data augmentation module in the pre-training (see \ref{subsec:method-contrastive}), data augmentations of random horizontal flip, random color jitter, and random grayscale conversion were used in the downstream tasks to mitigate the overfitting problem.
An ablation study on data augmentation is shown in Appendix C.
All images were resized to 224$\times$224, and the RGB channels were normalized using the mean and standard deviation of ImageNet~\cite{deng2009imagenet}.
In addition, for the weight prediction in the downstream task (not pre-training), each patient's weight was normalized to mean 0 and variance 1.
The predicted weights were then transformed back to the original scale using the mean and variance of the training data.

To verify the effectiveness of our WeightSupMoCo, we compared it to several methods (see Table~\ref{table:performance}). 
In all comparative methods, we used the same CNN model ({\it i.e.}, ResNet18) and data augmentation module as the proposed method.
Performance comparisons using commonly used CNN networks other than ResNet18 are shown in Appendix D.
To verify the effectiveness of the contrastive multi-patient pre-training, we compared our method with a model trained from scratch using only per-patient data ({\it i.e.}, without multi-patient pre-training).
To verify the effectiveness of using supervised labels ({\it i.e.}, pre- and post-dialysis labels and patient weight), we compared our representation learning with the self-supervised learning methods, SimCLR~\cite{chen2020simple}, MoCo~\cite{he2020momentum}, and SimSiam~\cite{chen2021exploring}.
SimSiam is one of the state-of-the-art self-supervised learning methods in computer vision.
We also compared it with standard supervised pre-training methods, classification using pre- and post-dialysis labels based on the cross-entropy loss and weight prediction based on the MSE loss.
In the weight prediction-based pre-training method, all patients' weights were normalized.
Furthermore, we compared our WeightSupMoCo loss with the SupCon loss~\cite{khosla2020supervised} using pre- and post-dialysis labels and the y-Aware InfoNCE loss~\cite{dufumier2021contrastive} using patient weight.
In addition, to verify the effectiveness of using MoCo-based contrastive learning in our framework, we compared WeightSupMoCo loss with a SimCLR-based WeightSupCon loss ({\it i.e.}, a SimCLR-based version of WeightSupMoCo). 
For a fair comparison, the downstream task settings were the same for all methods.
In the pre-training, the learning rates of cross-entropy and MSE losses were set to 1e-3, while the other learning rates and epoch numbers were the same as in our method.
The batch size was set to 16 for all methods.
The temperature parameter $\tau$ in contrastive learning SimCLR, MoCo, SupCon, {\it y}-Aware InfoNCE, WeightSupCon, and WeightSupMoCo was set to 0.1.
The dictionary size and momentum coefficient of MoCo and WeightSupMoCo were set to 1024 and 0.9999.
The hyperparameter of the RBF kernel $\sigma$ in {\it y}-Aware InfoNCE, WeightSupCon, and WeightSupMoCo was set to 3.0.

As evaluation metrics for the pre- and post-dialysis classification, we used accuracy, area under the receiver operating characteristic curve (ROC-AUC), and area under the precision-recall curve (PR-AUC).
As evaluation metrics for the weight prediction, we used mean absolute error (MAE), root mean squared error (RMSE), and correlation coefficient (CorrCoef) between the predicted and ground-truth data.

\begin{figure}[t]
\begin{center}
\includegraphics[scale=0.55]{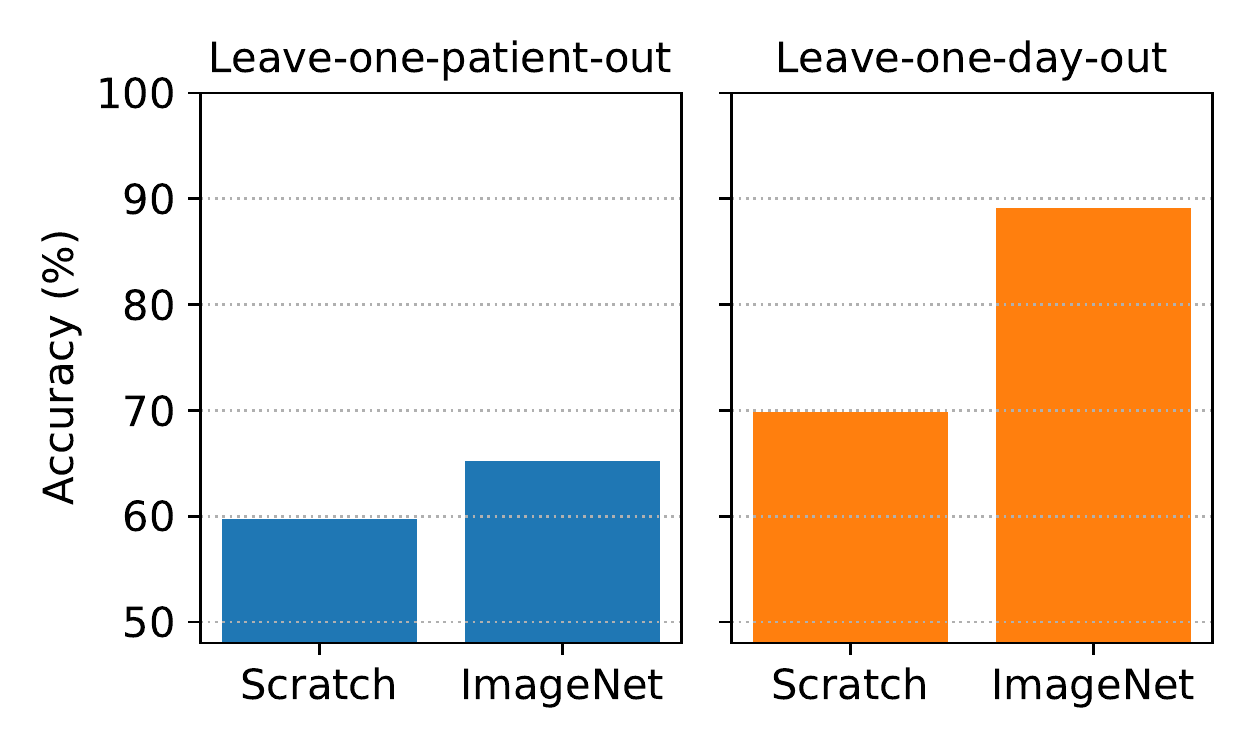}
\end{center}
\vspace{-15pt}
\caption{Accuracy of pre- and post-dialysis classification with leave-one-patient-out and leave-one-day-out cross-validations. A model trained from scratch and ImageNet-supervised pre-training model are compared.}
\label{fig:accuracy}
\end{figure}

\begin{table*}[t]
\caption{Estimation performance of pre- and post-dialysis classification and body weight prediction. Mean $\pm$ standard deviation of the estimation performance over 15 patients is shown.}
\vspace{-5pt}
\label{table:performance}
\begin{center}
\scalebox{0.92}{
\begin{tabular}{l|ccc|ccc}  \hline
\multicolumn{1}{c|}{} & \multicolumn{3}{c|}{Pre- and post-dialysis classification} & \multicolumn{3}{c}{Body weight prediction} 
\\ 
\cline{2-7}
Method &  Accuracy & ROC-AUC & PR-AUC & MAE & RMSE & CorrCoef \\ \hline\hline
Scratch &  69.8 $\pm$ 9.7 & 77.3 $\pm$ 11.9 & 78.3 $\pm$ 11.1 & 0.798 $\pm$ 0.194 & 0.934 $\pm$ 0.233 & 0.406 $\pm$ 0.234 \\ \hline
\textit{Self-supervised pre-training:} & & & & & \\
SimCLR~\cite{chen2020simple} & 75.3 $\pm$ 8.8 & 84.5 $\pm$ 9.4 & 84.7 $\pm$ 8.8 & 0.698 $\pm$ 0.204 & 0.835 $\pm$ 0.234 & 0.554 $\pm$ 0.179 \\
MoCo~\cite{he2020momentum} & 77.2 $\pm$ 9.2 & 86.3 $\pm$ 10.0 & 87.5 $\pm$ 9.0 & 0.648 $\pm$ 0.183 & 0.772 $\pm$ 0.211 & 0.612 $\pm$ 0.165 \\ 
SimSiam~\cite{chen2021exploring} & 73.3 $\pm$ 8.5 & 81.6 $\pm$ 10.3 & 82.5 $\pm$ 9.4 & 0.760 $\pm$ 0.183 & 0.899 $\pm$ 0.216 & 0.487 $\pm$ 0.185 \\ \hline
\textit{Supervised pre-training:} & & & & & \\
Classification (cross-entropy) & 80.2 $\pm$ 7.8 & 88.1 $\pm$ 7.5 & 88.7 $\pm$ 6.5 & 0.621 $\pm$ 0.161 & 0.737 $\pm$ 0.181 & 0.662 $\pm$ 0.124 \\ 
Weight prediction (MSE) & 82.4 $\pm$ 6.2 & 90.8 $\pm$ 7.4 & 91.9 $\pm$ 6.3 & 0.577 $\pm$ 0.186 & 0.706 $\pm$ 0.208 & 0.675 $\pm$ 0.139 \\
SupCon~\cite{khosla2020supervised} & 77.7 $\pm$ 9.4 & 86.6 $\pm$ 10.4 & 87.5 $\pm$ 9.6 & 0.665 $\pm$ 0.166 & 0.779 $\pm$ 0.186 & 0.599 $\pm$ 0.154 \\ 
y-Aware InfoNCE~\cite{dufumier2021contrastive} & 82.7 $\pm$ 7.2 & 91.7 $\pm$ 7.0 & 92.2 $\pm$ 7.0 & 0.575 $\pm$ 0.186 & 0.682 $\pm$ 0.200 & 0.699 $\pm$ 0.138  \\ 
WeightSupCon (SimCLR-based) & 84.7 $\pm$ 7.5 & {\bf 92.5 $\pm$ 7.6} & 92.9 $\pm$ 7.2 & 0.569 $\pm$ 0.182 & 0.676 $\pm$ 0.199 & 0.707 $\pm$ 0.135  \\ 
{\bf WeightSupMoCo (Ours)} & {\bf 84.9 $\pm$ 7.1} & {\bf 92.5 $\pm$ 7.6} & {\bf 93.0 $\pm$ 7.2} & {\bf 0.555 $\pm$ 0.184} & {\bf 0.626 $\pm$ 0.137} & {\bf 0.718 $\pm$ 0.136}  \\ \hline  
\end{tabular}
}
\end{center}
\end{table*}

\begin{table*}[t]
\caption{Estimation performance of pre- and post-dialysis classification and body weight prediction for patient-common model.}
\vspace{-5pt}
\label{table:performance-common}
\begin{center}
\scalebox{0.92}{
\begin{tabular}{l|ccc|ccc}  \hline
\multicolumn{1}{c|}{} & \multicolumn{3}{c|}{Pre- and post-dialysis classification} & \multicolumn{3}{c}{Body weight prediction} 
\\ 
\cline{2-7}
Method &  Accuracy & ROC-AUC & PR-AUC & MAE & RMSE & CorrCoef \\ \hline\hline
Patient-common model & 80.5 $\pm$ 6.4 & 88.5 $\pm$ 6.7 & 88.7 $\pm$ 6.2 &  0.986 $\pm$ 0.200 & 1.299 $\pm$ 0.294 & 0.394 $\pm$ 0.109 \\ \hline
\end{tabular}
}
\end{center}
\end{table*}

\subsection{Results}
\label{subsec:experiment-results}
\subsubsection{Generalizability to new patients not in training data}
First, we investigate the generalizability to new patients not in the training data when training a single model for all patients ({\it i.e.}, patient-common model), as discussed in \ref{sec:introduction}.
To evaluate the generalizability, we performed the pre- and post-dialysis classification with a {\it leave-one-patient-out cross-validation} using 39 patients, where the data from one patient were used as test data and the data from the other patients were used as training and validation data (the ratios of the training and validation data were 80\% and 20\%). 
To get an upper bound on the classification accuracy, we also show the averaged results when training models for each patient ({\it i.e.}, patient-specific model) with a leave-one-day-out cross-validation (see \ref{subsec:experiment-settings}) using 15 patients in the downstream task. 
Note that this experiment tests the generalizability of the patient-common model to new patients, and a fair comparison between the patient-common model and the patient-specific model is provided in \ref{subsec:experiment-results-2}. 
Figure~\ref{fig:accuracy} shows the accuracy of the pre- and post-dialysis classification with leave-one-patient-out and leave-one-day-out cross-validations.
As pre-training setups, we compare ResNet18 trained from scratch ({\it i.e.}, without pre-training) and ResNet18 initialized from ImageNet-supervised pre-training.
From the results of the leave-one-patient-out cross-validation, the classification accuracy reaches only 59.7\% for training from scratch and 65.2\% with ImageNet pre-training.
This low level of performance is because the changes in patients' faces before and after dialysis are subtle compared with the differences in faces between patients.
As a result, it is difficult to classify new patient faces into pre- and post-dialysis.
On the other hand, from the results of the leave-one-day-out cross-validation, the classification accuracy reaches 69.8\% for training from scratch and 89.1\% with ImageNet pre-training.
These results suggest that training models for each patient successfully identifies subtle changes in the face before and after dialysis.
We can thus classify facial images on a new day into pre- and post-dialysis relatively well.

\subsubsection{Effectiveness of WeightSupMoCo-based pre-training}
\label{subsec:experiment-results-2}
Table~\ref{table:performance} shows the estimation performances of the pre- and post-dialysis classification and body weight prediction of the proposed and comparative methods with the leave-one-day-out cross-validation.
Note that the models were not initialized with ImageNet pre-training in order to fairly evaluate the effectiveness of the representation learning of each pre-training method (comparisons with ImageNet pre-training are provided in \ref{subsec:experiment-results-3}).
From the table, it is clear that our WeightSupMoCo outperforms all of the comparative methods.
We investigated the statistical significance of WeightSupMoCo with a t-test using accuracy for pre- and post-dialysis classification and MAE for weight prediction. 
For the t-test, the results of all cross-validations were used. 
The significance was confirmed ($P < 0.05$) with all methods except WeightSupCon in the pre- and post-dialysis classification, and with all methods except Weight prediction (MSE) in the weight prediction.

In the following, we evaluate the effectiveness of each component from the pre-training methods.
First, a comparison between training from scratch and the other methods indicates the remarkable effectiveness of multi-patient pre-training.
Hence, pre-training with a larger dataset obtained from multiple patients can boost the estimation performance and thereby overcome the problem of the limited dataset obtainable from individual patients.
Next, a comparison of the results for the self-supervised pre-training methods ({\it i.e.}, SimCLR, MoCo, and SimSiam) with supervised pre-training methods shows the effectiveness of utilizing supervised labels about edema, {\it i.e.}, pre- and post-dialysis labels and patient weight.
Thus, we can successfully acquire knowledge about edema by using supervised labels in the multi-patient pre-training.
Specifically, by comparing Weight prediction (MSE) and y-Aware Info NCE with Classification (cross-entropy) and SupCon, we can see that patient weight is more effective than the pre- and post-dialysis labels.
This result suggests that the fine-grained label on edema may provide more meaningful information than the coarse-grained label.
Furthermore, by comparing WeightSupCon and WeightSupMoCo with Classification (cross-entropy), Weight prediction (MSE), SupCon, and y-Aware Info NCE, we can see that the use of both the pre- and post-dialysis labels and patient weight further enhances the estimation performance.
From this result, we also find that the use of both coarse- and fine-grained labels improves the feature representation.
Finally, by comparing WeightSupMoCo with WeightSupCon, we find that MoCo-based contrastive learning improves performance.
This is because our MoCo-based approach makes the dictionary size larger, and enables contrastive learning with a larger number of samples.

\begin{figure}[h]
\begin{center}
\includegraphics[scale=0.5]{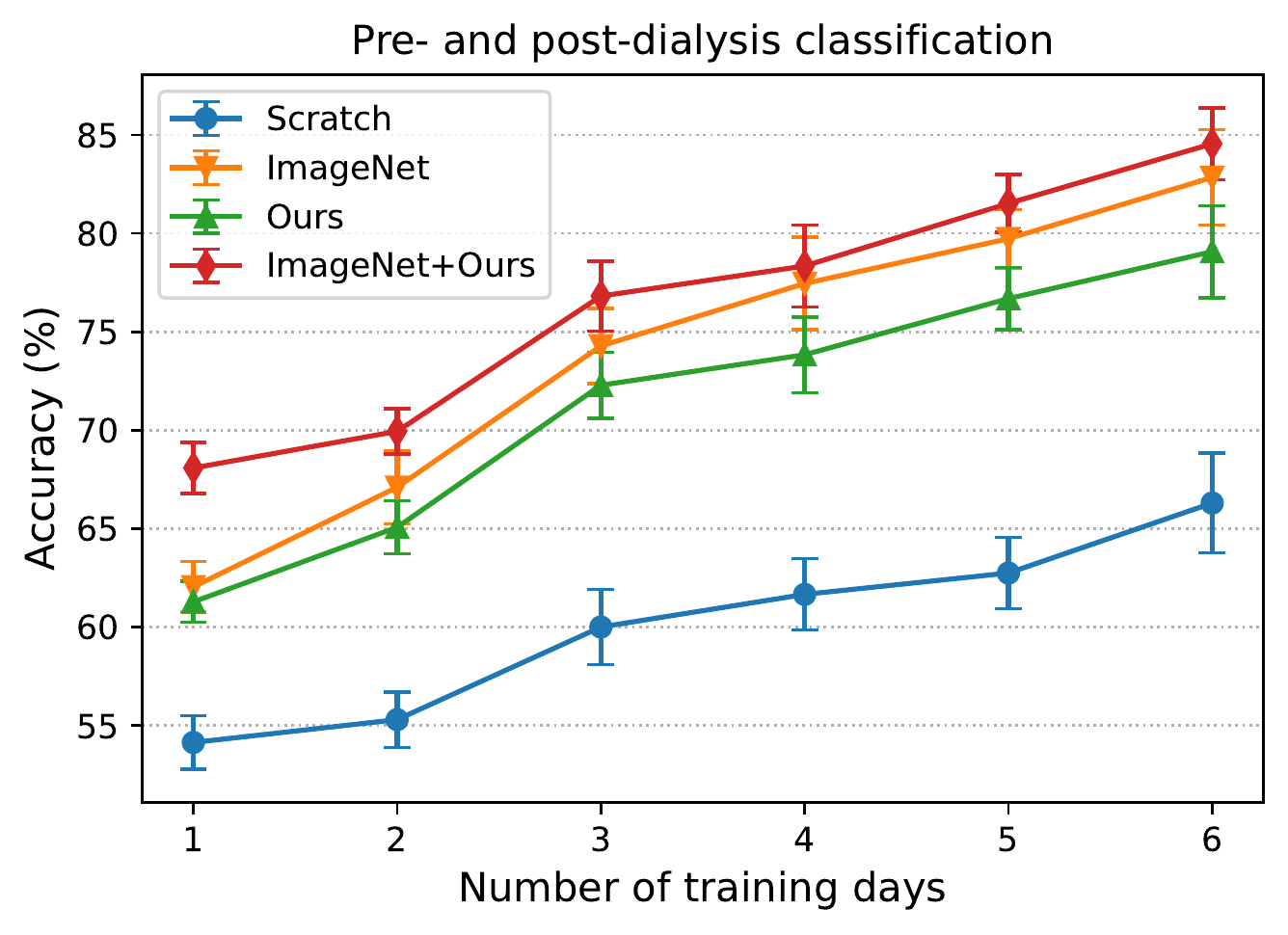}
\includegraphics[scale=0.5]{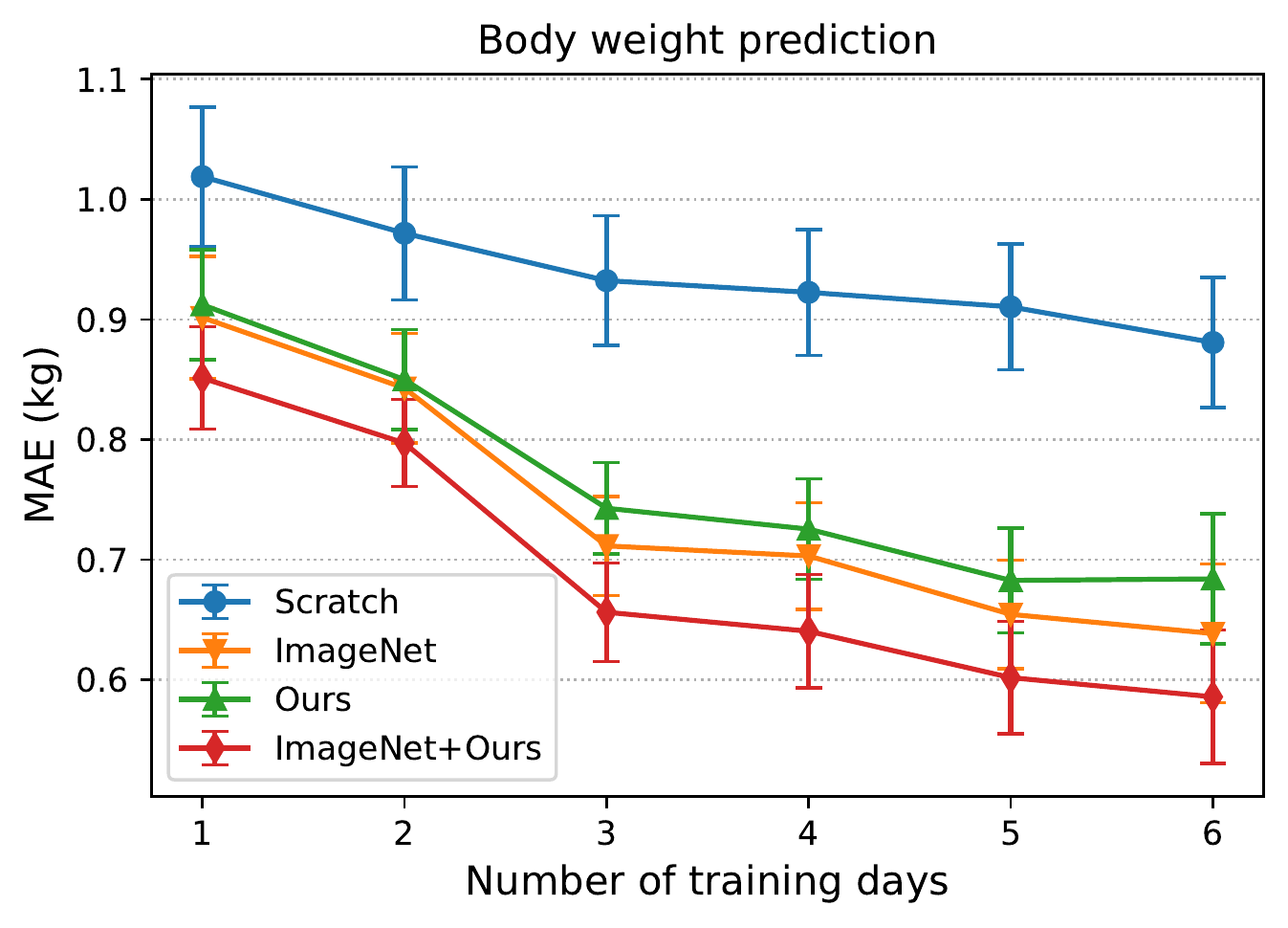}
\end{center}
\vspace{-5pt}
\caption{Relationship between the number of training days for transfer learning and estimation performance of pre- and post-dialysis classification (top) and body weight prediction (bottom). Error bars represent standard errors of all results for each training day.}
\label{fig:trnum}
\end{figure} 

We also show the estimation performance of the patient-common model with the leave-one-day-out cross-validation in Table~\ref{table:performance-common}. 
The model is not trained in two steps ({\it i.e.}, multi-patient pre-training and transfer learning for per-patient), but is trained with all patient data in a single step. 
The pre- and post-dialysis classification accuracy of the patient-common model is similar to its two-step version, cross-entropy. 
On the other hand, since the patient-common model is trained on multiple patients' weight data, the performance of weight prediction is much lower than our two-step training. 
WeightSupMoCo still outperforms the patient-common model.
Also, the training time of our model (transfer learning) is considerably shorter than that of the patient-common model, which is trained on all patients data. 
This is an important advantage because our method requires training before the patient can use it.

\subsubsection{Estimation performance vs. number of training days in downstream tasks}
\label{subsec:experiment-results-3}
Since the number of dialysis days in the training data ({\it i.e.}, training days) of each patient is limited, we aim to achieve more accurate estimation in the downstream tasks with fewer training days.
Figure~\ref{fig:trnum} shows the relationship between the number of training days for transfer learning and the estimation performance of pre- and post-dialysis classification and weight prediction.
When reducing the number of training days, dialysis days were selected randomly but these sets were identical for all methods.
For each training day, we randomly selected dialysis days five times for the training data and then averaged the performances across all trials.
As pre-training setups, we compare a model trained from scratch, ImageNet-supervised pre-training, WeightSupMoCo-based pre-training (Ours), and WeightSupMoCo-based pre-training initialized from ImageNet pre-training (ImageNet+Ours) models.
Also, when there are only a few training days, the model may be overfitted because of the small amount of training data.
Therefore, when there were only one or two training days, most of the layers were frozen and only part of them (the last conv5\_2 block~\cite{he2016deep} and the linear layer of ResNet18 were set after several trials) were trainable in ImageNet, Ours, and ImageNet+Ours.
When there were three to six training days, all layers of ResNet18 were trainable.
The estimation performances of all of the methods tend to improve as the number of training days are increased.
We also find that there are large performance differences between the scratch and the pre-training methods.
Among the pre-training methods, ImageNet is better than Ours, but this result may have heavily depended on the amount of pre-training data.
That is, ImageNet was pre-trained on 1.28 million images~\cite{he2016deep}, whereas Ours was pre-trained on only about 10,000 images, a difference in the number of images of more than 100 times.
Therefore, as the number of patients used in the pre-training increases in the future, Ours will likely show a performance improvement.
Furthermore, ImageNet+Ours boosts the estimation performance of ImageNet.
This is means that employing ImageNet-based and WeightSupMoCo-based pre-training together is effective.
Pre-training with ImageNet+Ours enables a relatively highly accurate estimation to be made with a small number of training days; {\it e.g.}, the accuracy is 76.8\% for the pre- and post-dialysis classification and MAE is 0.656 kg for the body weight prediction in the case of 3 training days.

\section{Discussion}
\label{sec:discussions}
We have demonstrated that the patient-specific model built using our pre-training strategy successfully performs pre- and post-dialysis classification and weight prediction from facial images.
In this section, we further validate the estimation results, and discuss how an edema estimation system based on the proposed method can be used in the future.
In the following, ImageNet+Ours refers to the proposed method that achieved the highest estimation performance in \ref{subsec:experiment-results}.

\begin{figure}[t]
\begin{center}
\includegraphics[scale=0.3]{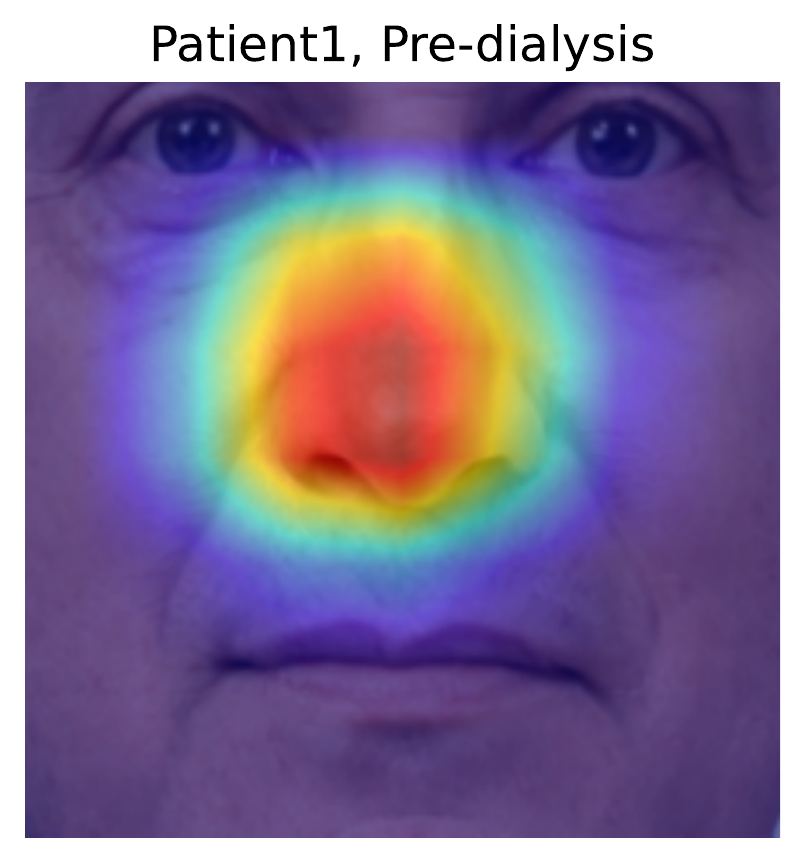}
\includegraphics[scale=0.3]{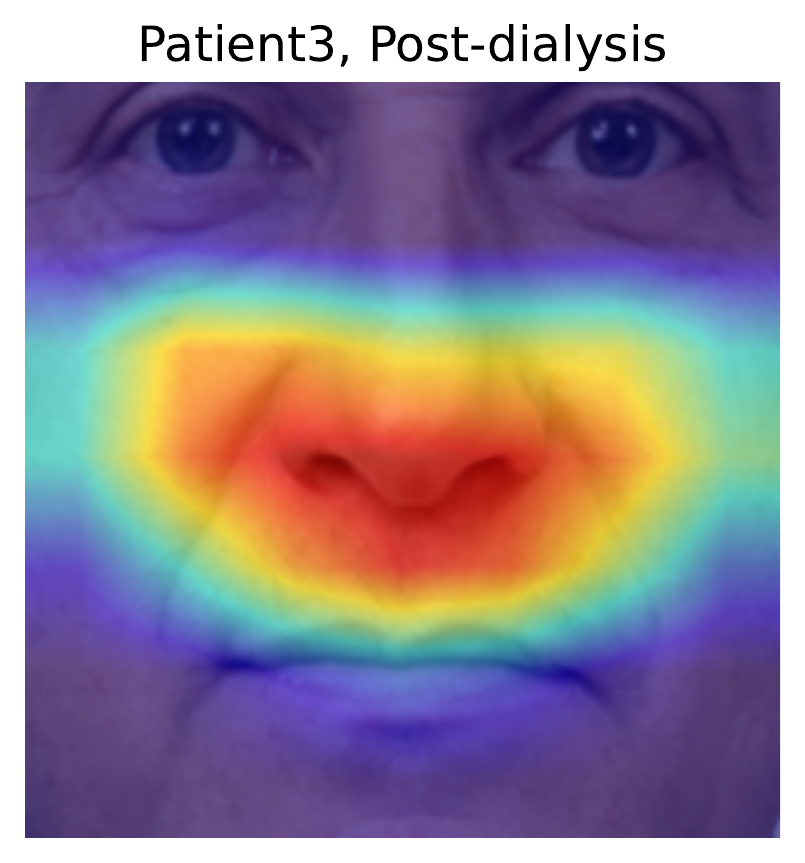}
\includegraphics[scale=0.3]{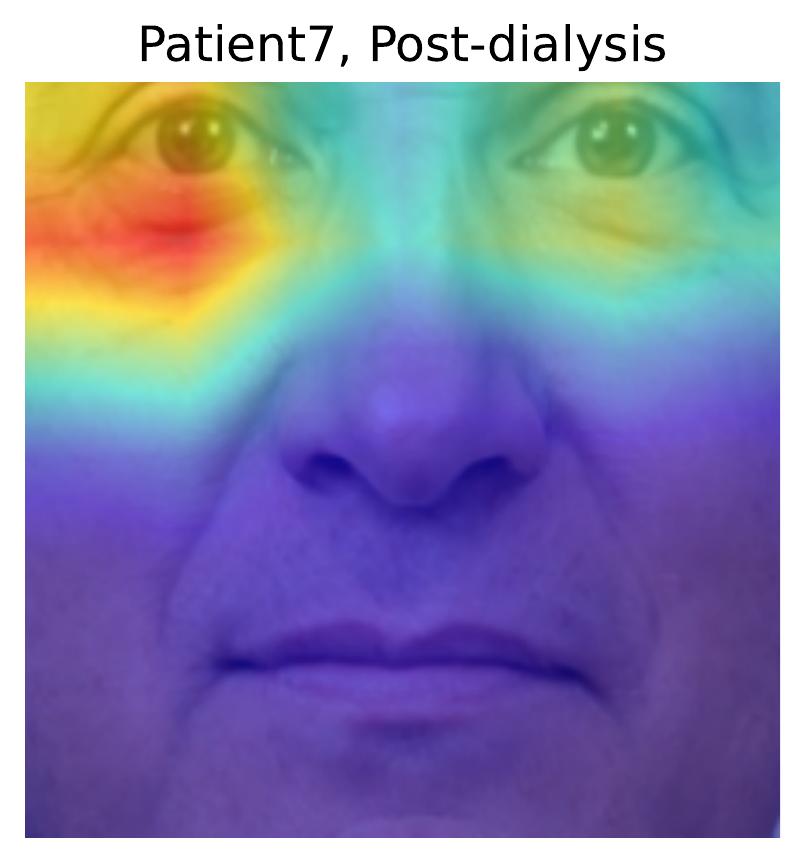}

\includegraphics[scale=0.3]{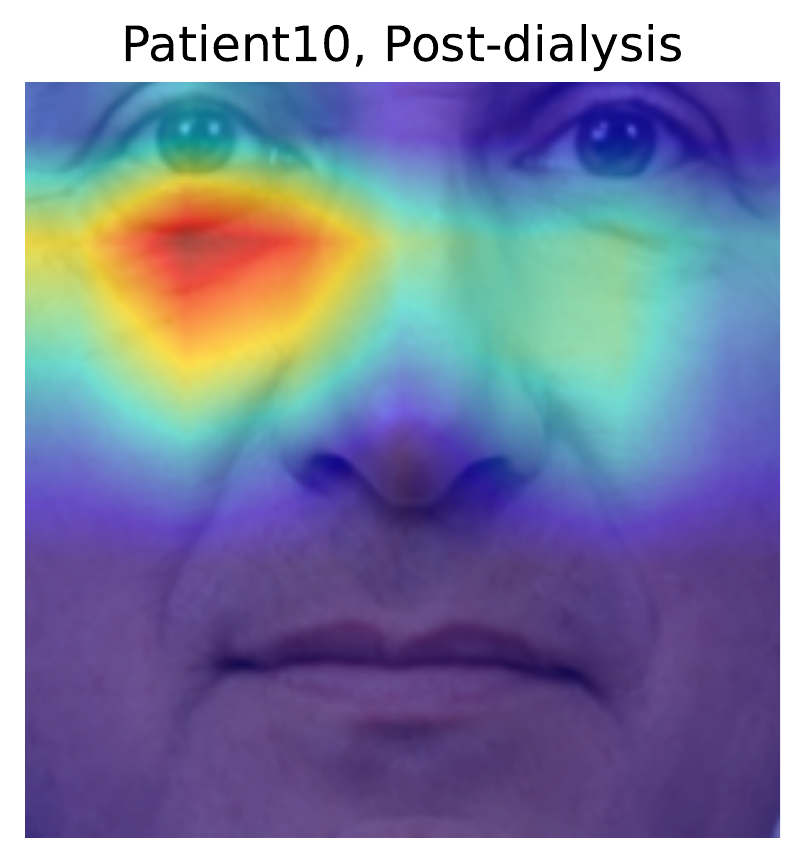}
\includegraphics[scale=0.3]{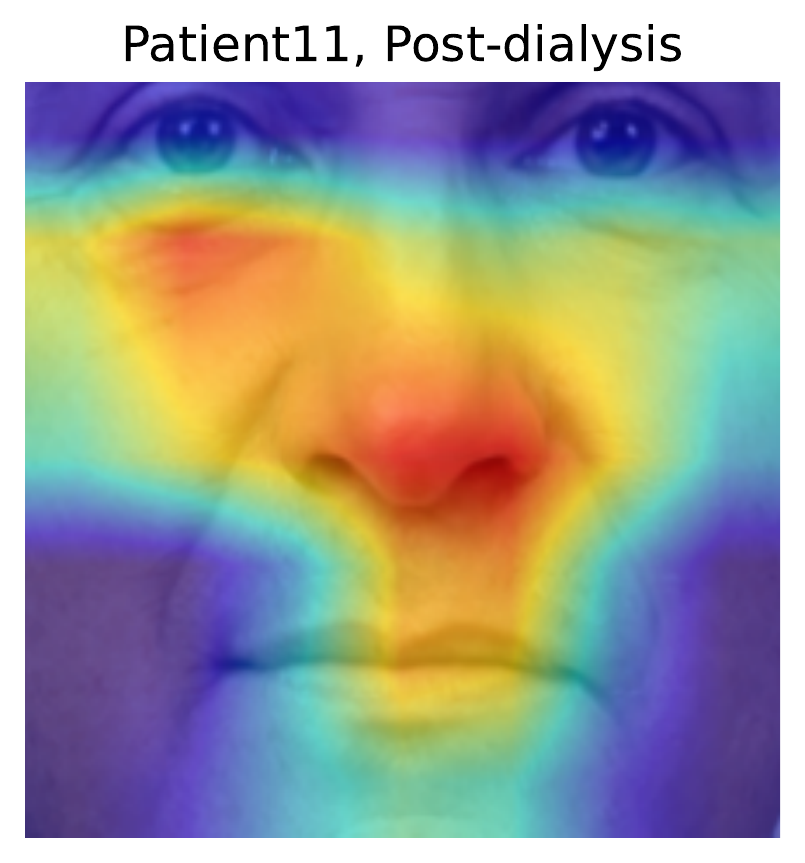}
\includegraphics[scale=0.3]{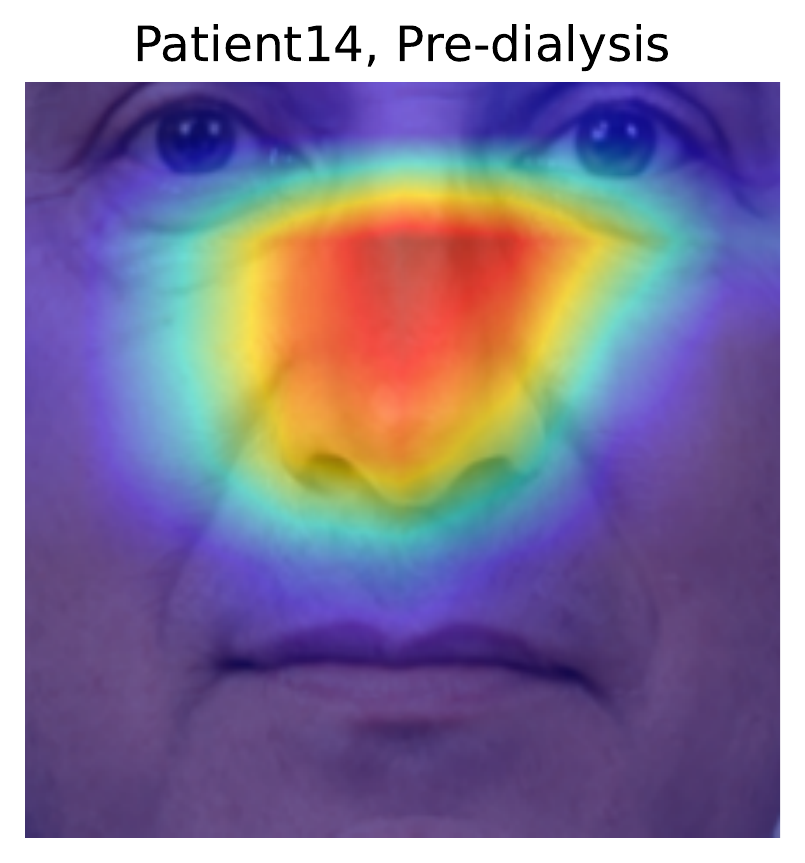}

\end{center}
\vspace{-10pt}
\caption{Visualization of saliency maps estimated by Grad-CAM when the proposed method classifies test facial images into pre- and post-dialysis. Above the image, the patient index and prediction results (consistent with ground-truth) for pre- and post-dialysis classification are shown.}
\label{fig:saliency}
\end{figure}

\begin{figure}[t]
\begin{center}
\includegraphics[scale=0.5]{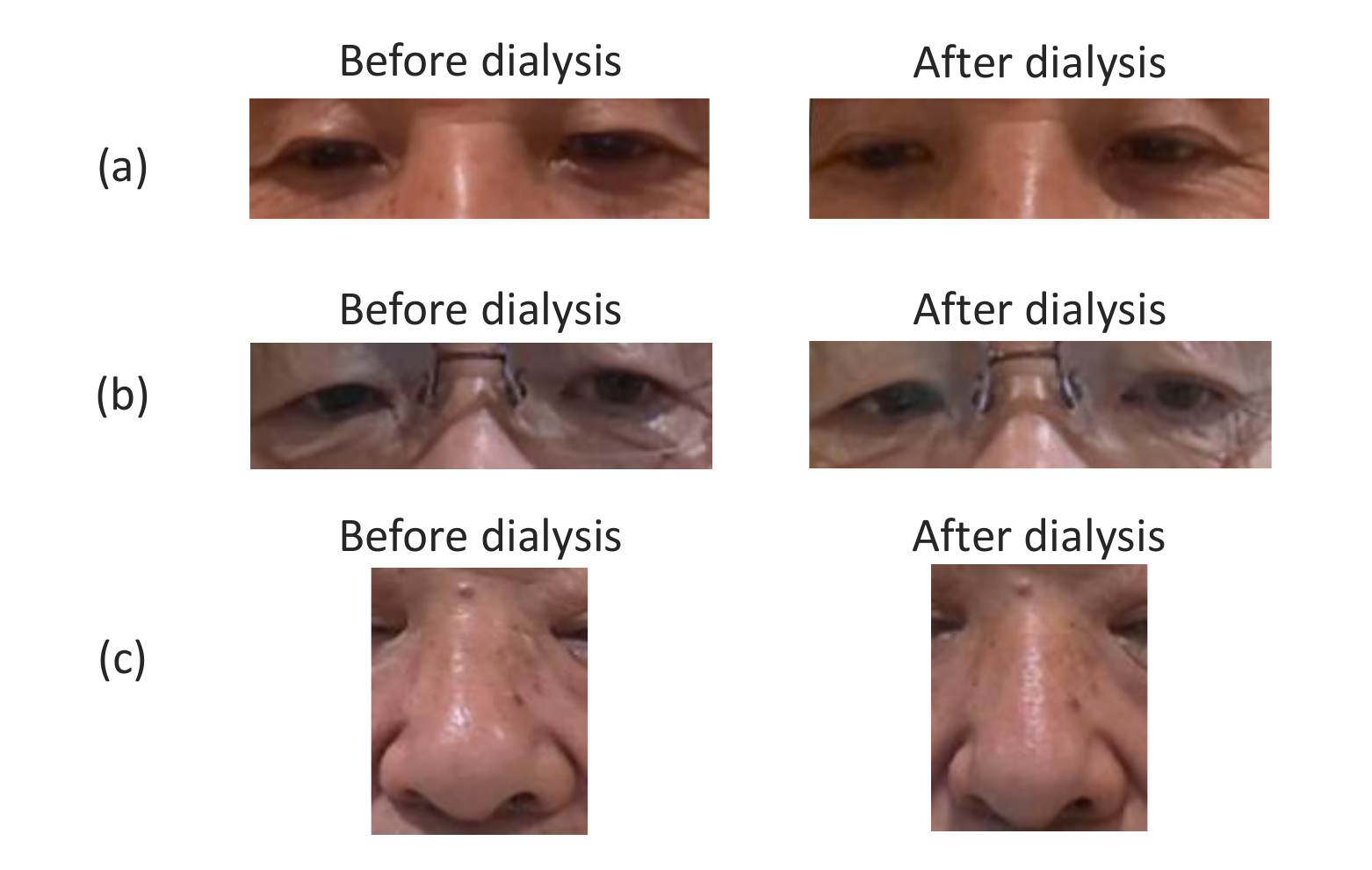}
\end{center}
\vspace{-15pt}
\caption{Part of actual patient's facial image before and after dialysis.}
\label{fig:actual}
\end{figure}

To understand the facial parts our model captures, we examine the visualization results of Grad-CAM~\cite{selvaraju2017grad}.
Figure~\ref{fig:saliency} shows~\footnote{Although the saliency maps are estimated from actual patients' facial images, the background facial images are not of the patients' faces but are Generated Photos (\url{https://generated.photos/}) to protect personal information.} the saliency maps estimated by Grad-CAM when the proposed method classifies test facial images into pre- and post-dialysis.
Here, we can see that our model uses mainly the eyes and nose for classification.
Figure~\ref{fig:actual} shows the actual facial images of patients when our model focused on the eyes and nose for classification. 
As shown in Fig.~\ref{fig:actual} (a) and (b), the appearance of the eyelids changed before and after dialysis ({\it e.g.}, edema-induced swelling was reduced, changing from a single eyelid to a double eyelid) due to the decrease in body fluid volume caused by dialysis. 
As shown in Fig.~\ref{fig:actual}~(c), we also observed that the shape of the nose changed before and after dialysis ({\it e.g.}, the edema-induced rounded shape disappears). 
Fig.~\ref{fig:saliency} and Fig.~\ref{fig:actual} suggest that our model captures these changes in the eyes and nose before and after dialysis.

\begin{figure}[t]
\begin{center}
\includegraphics[scale=0.5]{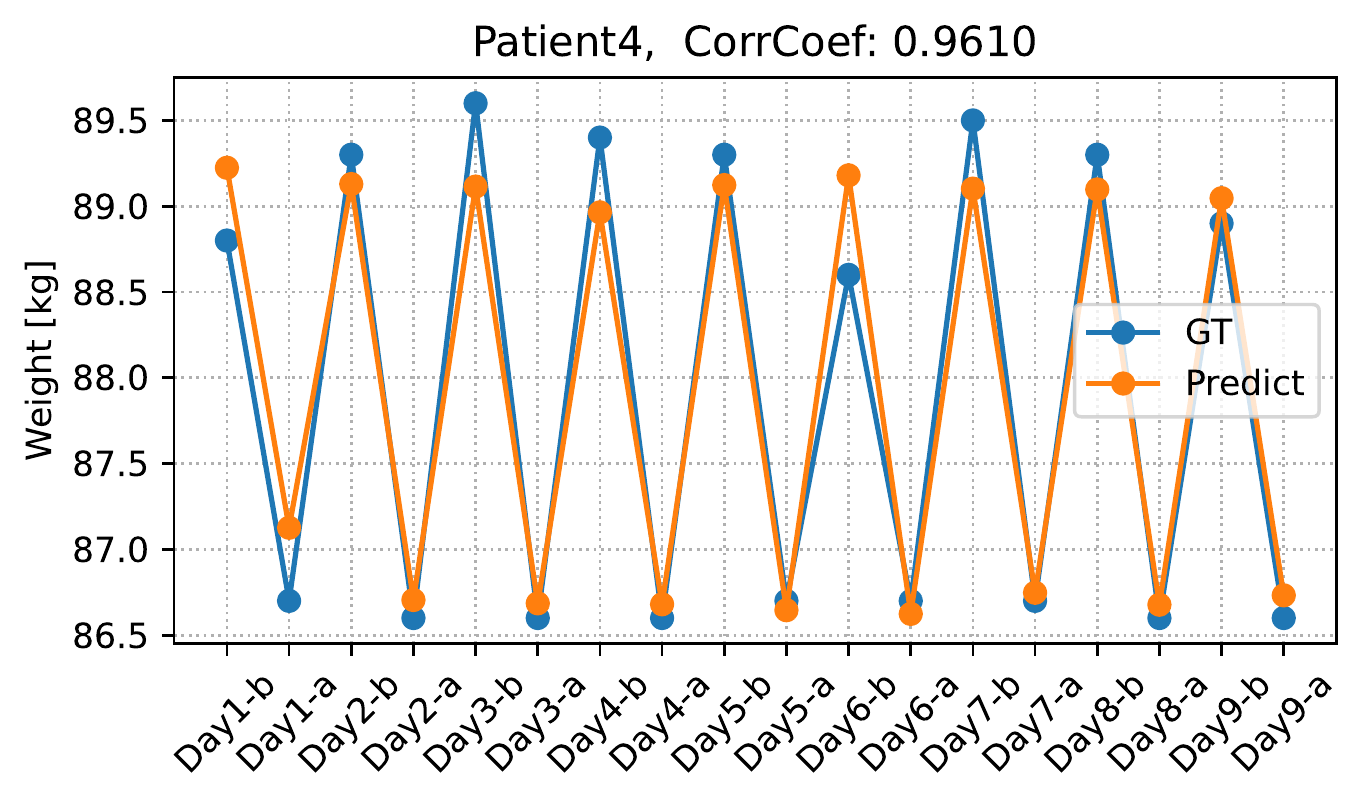}
\includegraphics[scale=0.5]{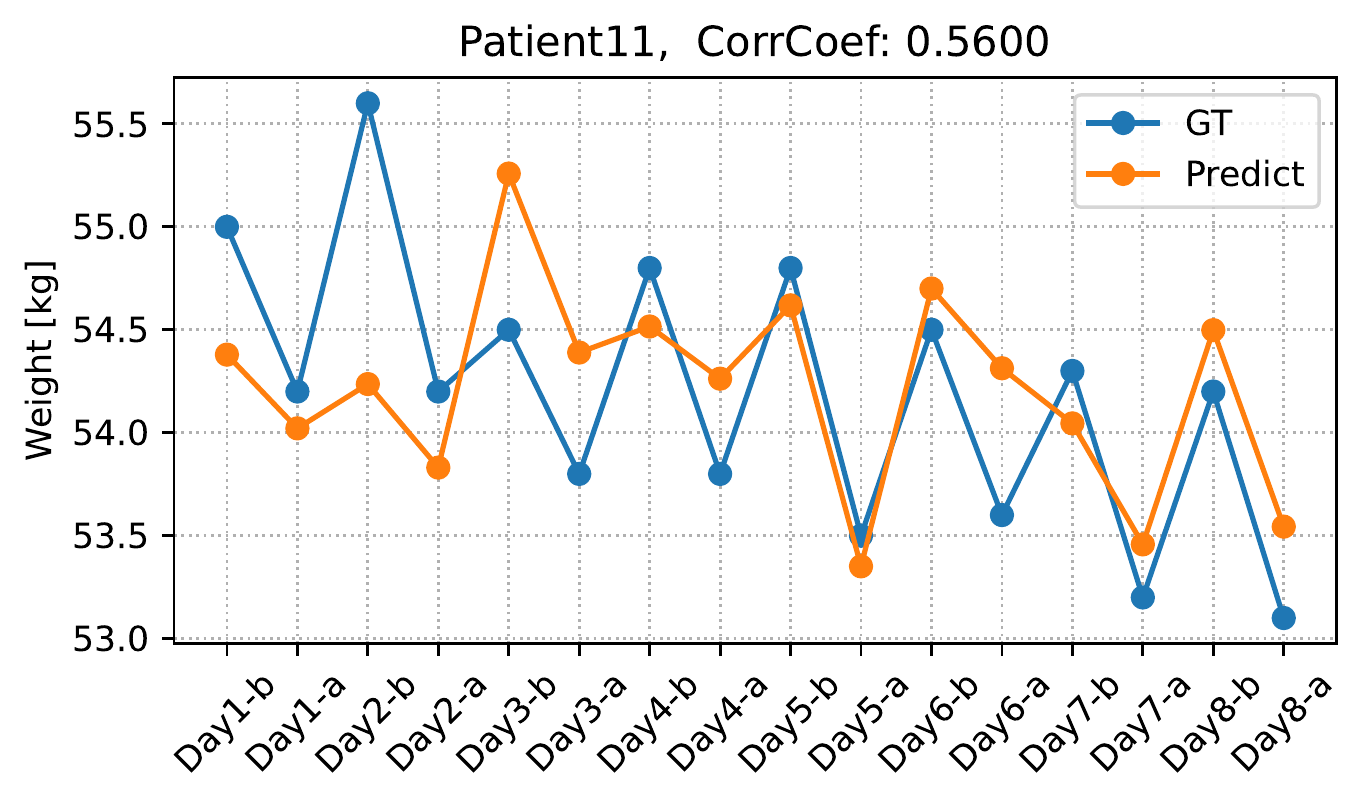}
\end{center}
\vspace{-15pt}
\caption{Changes in predicted and ground-truth (GT) body weights over several dialysis days (Top: the patient with the highest correlation coefficient, bottom: the patient with the lowest correlation coefficient). The horizontal line shows before (denoted by "b") and after (denoted by "a") dialysis, sorted by dates.}
\label{fig:weight}
\end{figure}

\begin{figure}[h]
\begin{center}
\includegraphics[scale=0.5]{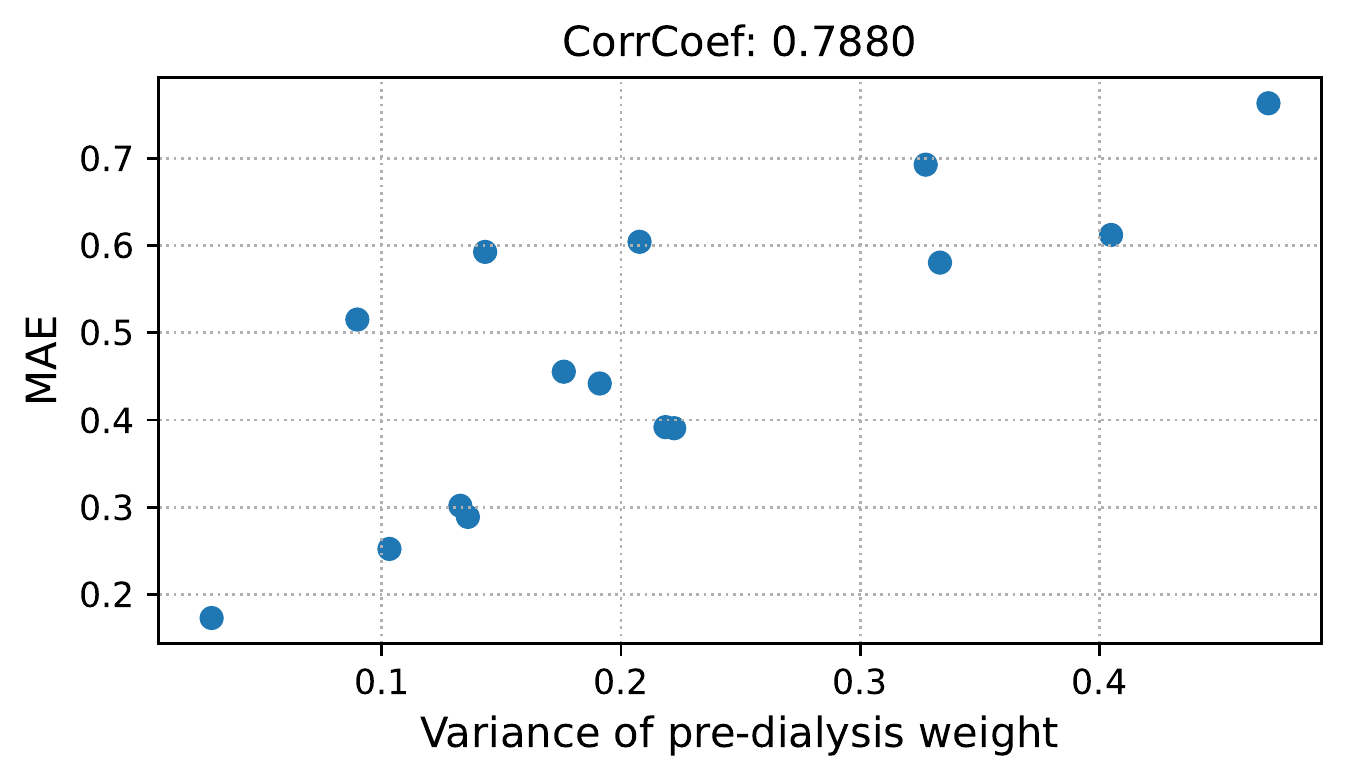}
\end{center}
\vspace{-15pt}
\caption{Relationship between the variance of pre-dialysis weights and the performance of weight prediction (MAE) in 15 patients.}
\label{fig:variance}
\end{figure}

\color{black}
We further verify the body weights predicted by the proposed method by conducting a leave-one-day-out cross-validation.
Figure~\ref{fig:weight} shows the changes in the predicted and ground-truth weights over all dialysis days.
The top figure is for the patient with the highest estimation performance, and the bottom figure is for the patient with the lowest estimation performance.
From the top figure, we see that our model can accurately predict weight changes throughout all dialysis days.
On the other hand, the bottom figure shows that our model may perform poorly on some dialysis days. 
Although the performance of weight prediction in day 2 and day 3 of patient 11 is low, the pre- and post-dialysis classification accuracy in these days is relatively high (82.4\% and 98.4\%, respectively). 
Therefore, our model can predict weight gain or loss before and after dialysis even though the performance of weight prediction is low. 
The reason for the poor performance of weight prediction can be due to day-to-day variation in weight. 
Figure~\ref{fig:variance} shows the relationship between the variance of pre-dialysis weight and MAE of weight prediction in 15 patients. 
Note that the variance is calculated only from pre-dialysis weights, since post-dialysis weights are determined approximately by pre-dialysis weights. 
The correlation coefficient is 0.7880, indicating a positive correlation between the variance of pre-dialysis weights and MAE. 
Thus, when the day-to-day variation in weight is large ({\it e.g.}, weight decreases from day 1 to day 8, as in patient 11), weight prediction becomes more difficult.
From the above results, even when we can classify facial images into pre- and post-dialysis, estimation of body weight, which is a proxy for the degree of edema, may be difficult in some cases.

A system using our edema estimation method could have several practical applications in the future. 
First, since our edema estimation system can easily estimate body weight from facial images, it has the potential to be a very powerful tool for dialysis patients who require strict weight control. 
While an accurate weight can be measured indoors on a weight meter, our system allows patients to obtain a reference weight anytime and anywhere by simply using the camera on their smartphone. 
For example, in a restaurant, patients can easily know how much fluid they should intake. 
Also, if our estimation performance becomes very accurate in the future, we may be able to solve the problem of weight errors due to clothing on weight meters.
In addition, the pre- and post-dialysis classification scores that our model predicts can allow patients to understand which of their conditions are similar to those before or after dialysis.
On the basis of the estimated weight and classification scores, dialysis patients can determine their food and fluid intake, meal menus, and amount of medicine ({\it e.g.}, diuretic).
The strength of this system is that users' existing tablets and smartphones can be used without any additional special equipment.

Although experimental results are promising, the proposed method still has some limitations.
First, in order to construct a patient-specific model, it requires training data from a patient who uses our system.
While our multi-patient pre-training strategy can reduce the number of training days needed to achieve high estimation performance (see Fig. \ref{fig:trnum}), a further reduction in training days is desired.
Second, this study examined only 39 Japanese patients, and our findings may not be generalizable to other racial groups.
In future work, we should increase the number of patients and validate our findings in other racial groups.
Third, this study used only facial images taken before and after dialysis and was not validated using facial images taken on non-dialysis days.
Considering that our edema estimation system will be used in the daily lives of dialysis patients, we should evaluate its estimation performance using facial images taken on non-dialysis days.

\section{Conclusions}
\label{sec:conclusions}
This study aimed to estimate the degree of edema from facial images taken before and after dialysis.
To estimate the degree of edema, we performed pre- and post-dialysis classification and weight prediction.
We developed a multi-patient pre-training strategy and transferred the pre-trained model to a patient-specific model.
To effectively pre-train the model with knowledge about edema from multiple patients, we have proposed a novel contrastive representation learning, called WeightSupMoCo.
Experiments conducted on facial images taken from 39 patients over a total of 210 dialysis days demonstrated the effectiveness of our approach.
The informative results of this study suggest that our edema estimation system could make a significant contribution to health monitoring of dialysis patients.

\section*{Acknowledgments}
The authors would like to thank Dr. Katsumi Takemura, director of the Takemura Medical Nephro Clinic, and the clinic's medical staff for their cooperation in patient recruitment and data collection.
The authors also would like to thank all patients who participated in this study.

\section*{Appendix A}
\begin{figure}[h]
\begin{center}
\includegraphics[scale=0.45]{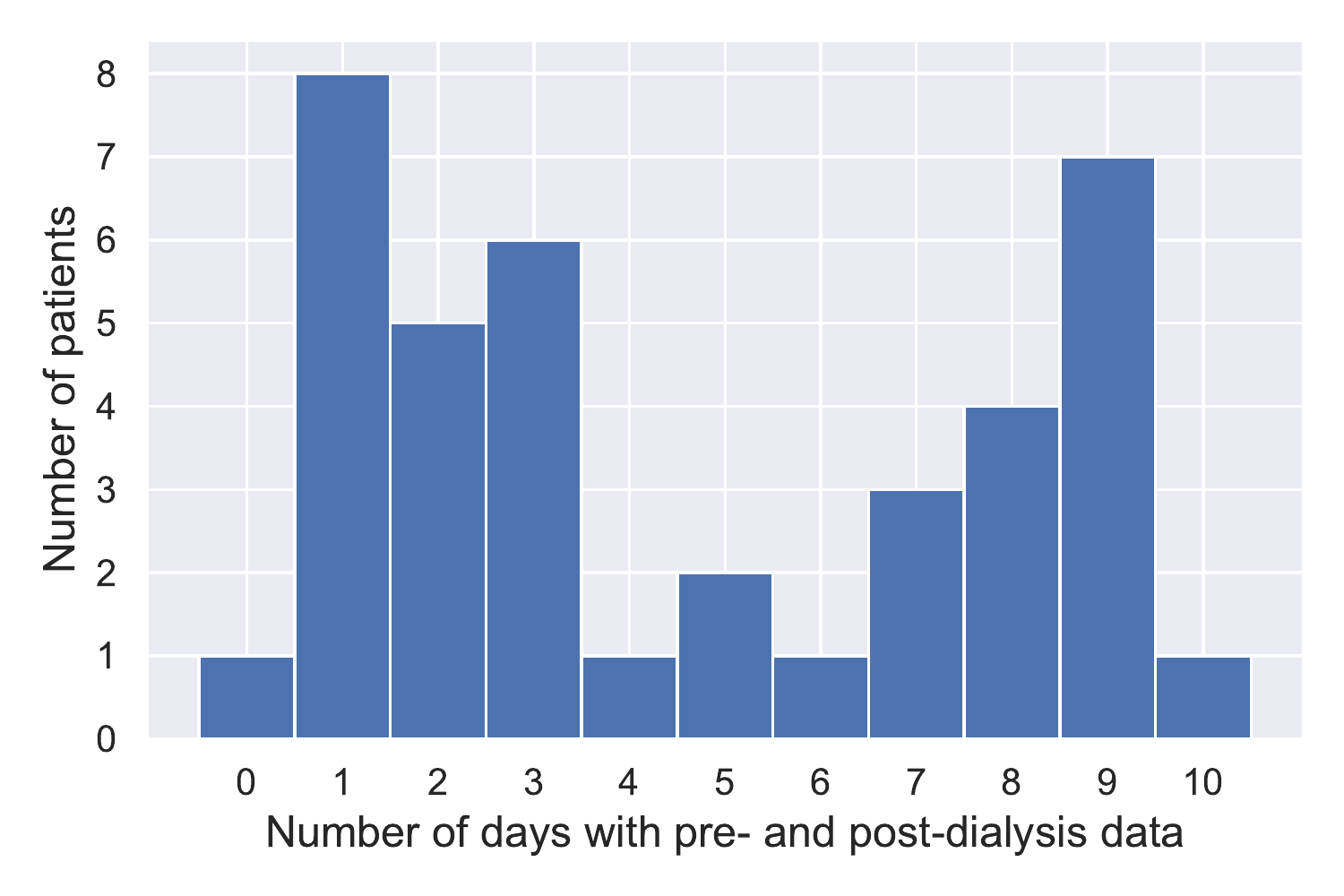}
\end{center}
\vspace{-15pt}
\caption{Relationship between the number of days and the number of patients.}
\label{fig:dataset}
\end{figure}

In our experiment, each patient participated in data collection for a different number of days. 
Figure~\ref{fig:dataset} shows the relationship between the number of days and the number of patients.
Note that the horizontal axis is the number of days including both pre- and post-dialysis data (0 days means that only pre-dialysis data are included).

\section*{Appendix B}
\begin{figure}[h]
\begin{center}
\includegraphics[scale=0.5]{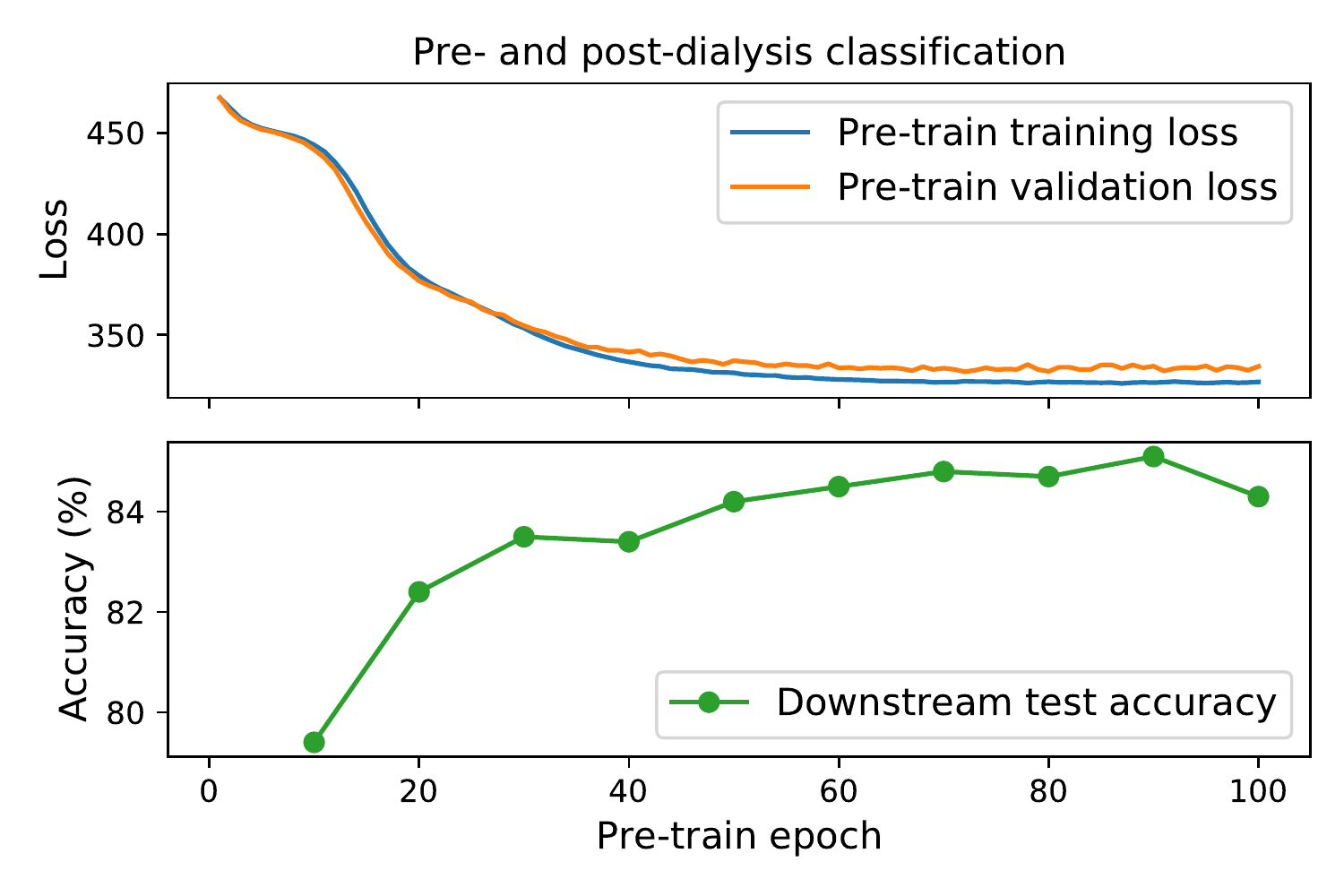}
\end{center}
\vspace{-15pt}
\caption{Training and validation losses in WeightSupMoCo pre-training and downstream test accuracy in pre- and post-dialysis classification.}
\label{fig:acc_epoch}
\end{figure}

In our multi-patient pre-training setting, the same patients were used for training and validation data to use as many patients as possible in the training data. 
However, the training and validation data are not split by patient, that can cause overfitting. 
Therefore, we monitor training and validation losses in the pre-training and downstream test accuracy in the pre- and post-dialysis classification. 
Figure~\ref{fig:acc_epoch} shows the training and validation losses in WeightSupMoCo pre-training from 1 to 100 epochs and the downstream test accuracy for every 10 epochs. 
We find that the test accuracy improves at the epochs where the validation loss converges. 
Therefore, monitoring the validation loss is sufficient for achieving good test performance in the downstream task.

\section*{Appendix C}
\label{sec:appen-c}
We use random horizontal flip, random color jitter, and random grayscale conversion as the data augmentation module in the multi-patient pre-training and downstream edema estimation. 
Here, we investigate the effectiveness of each data augmentation module in the WeightSupMoCo-based pre-training method. 
Table~\ref{table:aug} shows accuracy for the pre- and post-dialysis classification without data augmentation (w/o Aug), with each augmentation, and with a combination of all augmentation (Flip+Gray+Color Jitter) in the pre-training and downstream task. 
By comparing with w/o Aug, we confirm that any data augmentation is effective and combining all data augmentations further improves estimation performance. 
Among each data augmentation, changing the color of the image, such as color jitter and grayscale conversion, is particularly effective. 
This may be because the dataset size is small, causing overfitting to the lighting conditions of the original image. 
The data augmentation of changing colors can mitigate the overfitting and allow the model to focus on capturing edema features.

\begin{table}[h]
\caption{Ablation study on data augmentation. Accuracy for pre- and post-dialysis classification is shown.}
\vspace{-5pt}
\label{table:aug}
\begin{center}
\scalebox{0.75}{
\begin{tabular}{l|ccccc}  \hline
 & w/o Aug & Flip & Gray & Color Jitter & Flip+Gray+Color Jitter \\ \hline \hline
Accuracy & 72.1 & 75.1 & 76.9 &80.5 & 84.9 \\ \hline
\end{tabular}
}
\end{center}
\end{table}

\section*{Appendix D}
\label{sec:appen-d}
\begin{figure}[h]
\begin{center}
\includegraphics[scale=0.5]{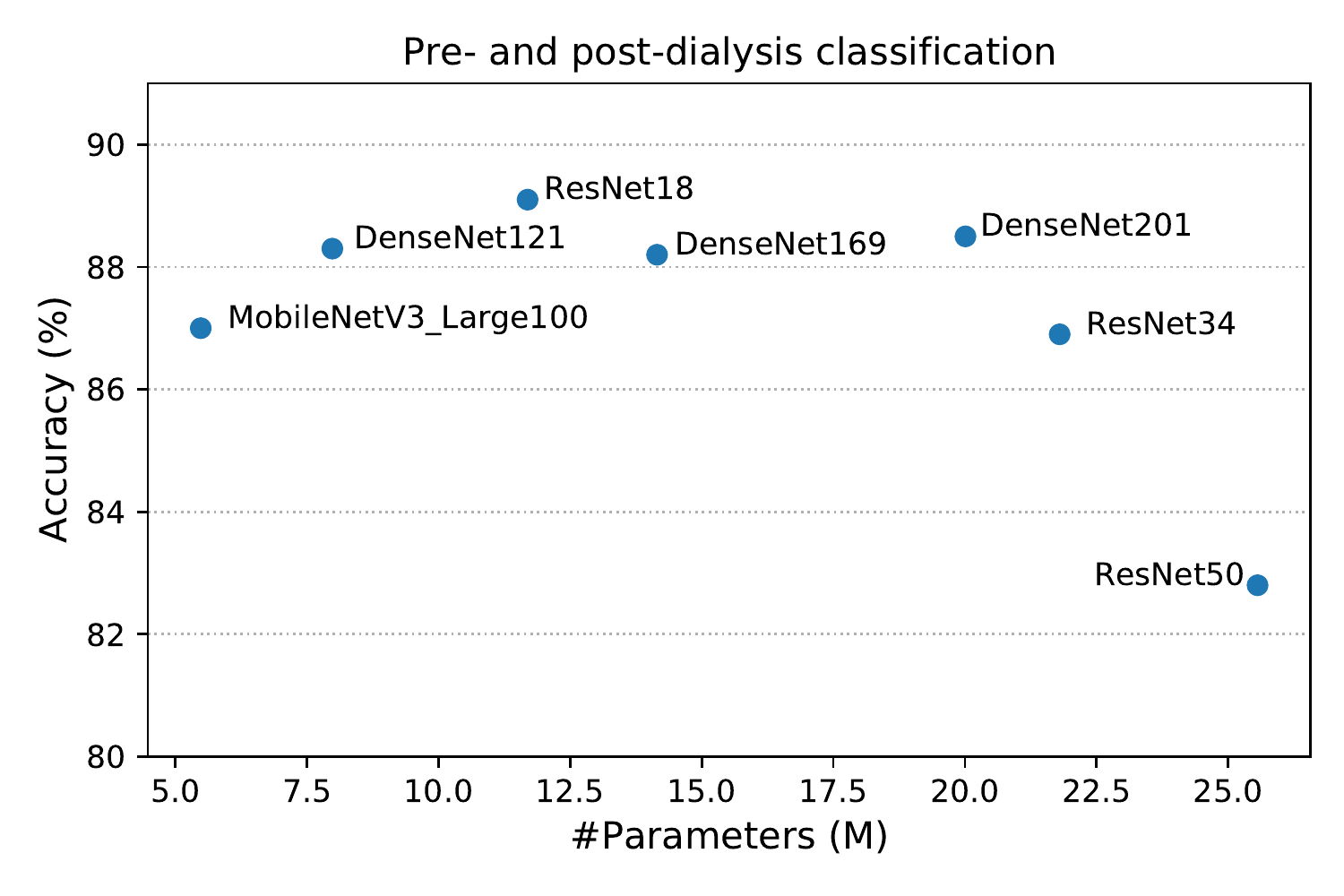}
\end{center}
\vspace{-15pt}
\caption{Estimation performance of commonly used CNN networks.}
\label{fig:acc_model}
\end{figure}

In all our experiments, we use ResNet18 as the backbone. Here, we show the estimation performance of other commonly used networks. 
Figure~\ref{fig:acc_model}  shows the pre- and post-dialysis classification accuracy of several networks initialized with ImageNet-supervised pre-training. 
We find that ResNet18 achieves the highest accuracy compared with other networks. 
ResNet18 is better than ResNet34 and ResNet50, which have a larger number of parameters. 
This could be because our dataset size is small and a small-scale network performs better.

\bibliographystyle{IEEEbib}
\bibliography{main}

\end{document}